%% file: mass-mapping.tex
\documentclass{emulateapj}
\usepackage[OT1]{fontenc}
\usepackage{xspace} 

\usepackage[pdfpagemode={UseOutlines},backref,breaklinks,bookmarks,bookmarksopen,colorlinks,linkcolor={black},citecolor={blue},urlcolor={red}]{hyperref}
\usepackage[all]{hypcap}

\usepackage{bm}
\usepackage{amsmath,amssymb}
\usepackage{aas_macros}
\usepackage{natbib}
\usepackage{bbold} 

\usepackage{multirow}
\usepackage[]{algorithm2e}

\let\oldtheequation\theequation
\makeatletter
\def\tagform@#1{\maketag@@@{\ignorespaces#1\unskip\@@italiccorr}}
\renewcommand{\theequation}{(\oldtheequation)}
\makeatother

\graphicspath{{figs-mass-mapping/}}

\usepackage{xkeyval}
\makeatletter
\newlength{\sfp@hseplen}\newlength{\sfp@vseplen}
\define@cmdkey{subfigpos}[sfp@]{pos}[ul]{}
\define@cmdkey{subfigpos}[sfp@]{font}[\small]{}
\define@cmdkey{subfigpos}[sfp@]{vsep}[2\baselineskip]{\setlength{\sfp@vseplen}{\sfp@vsep}}
\define@cmdkey{subfigpos}[sfp@]{hsep}[10pt]{\setlength{\sfp@hseplen}{\sfp@hsep}}
\newcommand{\subfigimg}[3][,]{%
  \setkeys{Gin,subfigpos}{pos,font,vsep,hsep,#1}
  \setbox1=\hbox{\includegraphics{#3}}
  \ifnum\pdfstrcmp{\sfp@pos}{ul}=0
    \leavevmode\rlap{\usebox1}
    \rlap{\hspace*{\sfp@hsep}\raisebox{\dimexpr\ht1-\sfp@vsep}{\sfp@font{#2}}}
    \phantom{\usebox1}
  \else\ifnum\pdfstrcmp{\sfp@pos}{ur}=0
    \leavevmode\usebox1
    \llap{\raisebox{\dimexpr\ht1-\sfp@vsep}{\sfp@font{#2}}\hspace*{\sfp@hsep}}
  \else\ifnum\pdfstrcmp{\sfp@pos}{lr}=0
    \leavevmode\usebox1
    \llap{\raisebox{\sfp@vsep}{\sfp@font{#2}}\hspace*{\sfp@hsep}}
  \else
    \leavevmode\rlap{\usebox1}
    \rlap{\hspace*{\sfp@hseplen}\raisebox{\sfp@vsep}{\sfp@font{#2}}}
    \phantom{\usebox1}
  \fi\fi\fi
}
\makeatother

\input{mbi-general-macros.tex}
\input{mbi-notation-macros.tex}

\input{mbi-addresses.tex}

\newcommand{\Cov}{\mathrm{Cov}}
\newcommand{\Matrix}[1]{\mathsf{#1}}

\begin{document}

\slugcomment{{LLNL-JRNL-697380-DRAFT}}

\title{Probabilistic cosmological mass mapping from weak lensing shear}
\shortauthors{Schneider et al.}

\author{M.~D.~Schneider\altaffilmark{1},
K.~Y.~Ng\altaffilmark{2,3},
W.~A.~Dawson\altaffilmark{1},
P.~J.~Marshall\altaffilmark{4,5},
J.~Meyers\altaffilmark{4},
D.~J.~Bard\altaffilmark{3}
}

\altaffiltext{1}{\llnl}
\altaffiltext{2}{\davis}
\altaffiltext{3}{\NERSC}
\altaffiltext{4}{\kipac}
\altaffiltext{5}{\slac}


\date{Draft \today\\\vspace{0.04in}}

\begin{abstract}
We infer gravitational lensing shear and convergence fields from galaxy ellipticity
catalogs under a spatial process prior for the lensing potential.
We demonstrate the performance of our algorithm with simulated Gaussian-distributed
cosmological lensing shear maps and a reconstruction of the mass distribution of the merging galaxy
cluster Abell 781 using galaxy ellipticities measured with the Deep Lens Survey.
Given interim posterior samples of lensing shear or convergence fields on the sky, we describe an
algorithm to infer cosmological parameters via lens field marginalization.
In the most general formulation of our algorithm we make no assumptions about weak shear or
Gaussian distributed shape noise or shears.
Because we require solutions and matrix determinants of a linear system of dimension that scales
with the number of galaxies, we expect our algorithm to require parallel high-performance
computing resources for application to ongoing wide field lensing surveys.
\end{abstract}


\section{Introduction} 
\label{sec:introduction}

The state of the inhomogeneous cosmic matter distribution (and the time-dependent
dark energy density) can be measured by
the gravitational lensing shearing and magnification of background luminous
sources. Using galaxies as the back lights for gravitational lensing (``cosmic shear'')
allows tomographic reconstruction of the matter distribution along the line of sight.
But, inference of galaxy lensing shears is clouded by the unknown distribution
(and possibly correlation) of galaxy intrinsic morphologies as well as image systematics
that induce spurious ellipticity correlations.

Common inference algorithms for gravitational lensing shear involve cross-correlating
the ellipticities of galaxies in a two-point function estimator under the assumption
of calibrated intrinsic ellipticity distributions and alignments~\citep[e.g.,][]{jee2015}.
This is a lossy procedure because the angular phase information in the lensing
shear and magnification fields on the sky is discarded. Traditional algorithms
are also necessarily biased because of the need to calibrate the unknown galaxy
ellipticity distributions. Said another way, the two-point function summary statistics of
cosmological large-scale structure do not capture the full statistical information in the multivariate
distribution of the lensing observables
\citep[e.g.,][]{pan2005,hamilton2006,takada2013,carron2014,2016arXiv160501100P}

In this paper we revisit the problem of inferring the lensing convergence posterior distribution
from a catalog of galaxy ellipticities to potentially capture more cosmological information from
cosmic shear than is available in two-point function estimators of the shear. We pursue a
probabilistic approach to lensing convergence inference as a means to propagate the large
uncertainties that can come in the presence of galaxy shape noise, finite survey areas, masking,
and sample selection. Moreover, we aim to build an inference framework that can fit within our previous
work on probabilistic cosmic shear~\citep{mbi-theory}.

\citet{1990ApJ...349L...1T} first presented a method to reconstruct the projected mass distribution
from weak gravitational lensing shears of source galaxy images.
\citet{1993ApJ...404..441K} derived the lensing convergence in terms
of the shear (a nonlocal relation) and applied this theory to estimators of the convergence given a
measured galaxy ellipticity catalog. The method of \citet{1993ApJ...404..441K}, while theoretically sound,
requires unbroken sky coverage and low-pass filtering to yield finite noise in the lensing
convergence estimator.
The \citet{1993ApJ...404..441K} method remains useful, however, for visualization
purposes~\citep{2015PhRvL.115e1301C,2015PhRvD..92b2006V}.
Subsequent papers have extended, applied, and explored the limitations of filtering algorithms for
cluster mass mapping~\citep{1995A&A...303..643B,1995A&A...294..411S,1998A&A...335....1L,2002ApJ...568..141G,2005A&A...440..453D,2007Natur.445..286M,2012A&A...540A..34D,2012MNRAS.424..553A,2013MNRAS.433.3373V}. Remaining data analysis challenges
using such methods include finite survey boundaries and masks,
separation of E and B modes in the shear
field, noise or significance characterization for shear or convergence extrema, and the requirement
in many algorithms to smooth or average the ellipticities of galaxies before the convergence inference
process.

Maximum likelihood (ML) estimators for the lensing
convergence~\citep[e.g.,][]{1996ApJ...464L.115B,1998A&A...337..325S,2008MNRAS.385.1431H,2009ApJ...702..980K}
can help mitigate biases arising from survey masks and admit mathematically consistent noise
characterization~\citep{2000MNRAS.313..524V}. However such estimators are biased. And many
approaches often still require a preliminary smoothing of the observed galaxy ellipticity field.

\citet{bridle1998} and \citet{bridle2000} introduced a `maximum entropy'
Bayesian prior for the lensing convergence from information theoretic and Bayesian analysis perspectives to
derive an estimator for the projected mass distribution of galaxy clusters with desirable noise
properties in a finite field.
\citet{Marshall:531310} and \citet{2001A&A...368..730S} refined the maximum entropy method to
specify the optimal smoothing
length scale for galaxy cluster mass inference via the Bayesian evidence of the observed
ellipticities. \citet{2013ascl.soft08004M} released a code\footnote{\lensent, \url{http://www.slac.stanford.edu/~pjm/lensent/version2/index.html}}
implementing the algorithm from \citet{Marshall:531310}.
The algorithm of \citet{Marshall:531310} is close to meeting all the requirements for the current
analysis, except the choice of smoothing scale and application to field rather than cluster lensing
is not demonstrated in the literature. We will show further benefits of the algorithm developed
in this paper below.
\citet{2006A&A...451.1139S} applied a modification of the maximum entropy method to a cosmic
shear analysis using $N$-body cosmological simulations to create mock observations.
The work of \citet{2006A&A...451.1139S} and also \citet{jiao2011} include maximum entropy algorithm
modifications to better handle masking, shape noise, and degrees of smoothing.

For constraining cosmological parameters, several groups have considered the abundance of peaks
in lensing convergence maps~\citep{2010PhRvD..81d3519K,2011ApJ...735..119S,2014MNRAS.442.2534S,2015PhRvD..91f3507L,2016MNRAS.456..641R,2016PhRvD..94d3533L,PhysRevLett.117.051101,2016arXiv160501100P}. These studies have a common
approach in direct calculation or estimation of lensing peaks without an attempt to infer
associated peaks in the 3D cosmological mass density. The measurement process is thus direct as
lensing by multiple structures or voids along any given line of sight can easily confuse the
2D to 3D mass inference. Just as the abundance of galaxy clusters provides tight constraints on the
cosmological model, so too can the abundance of lensing peaks.
However, the estimators for lensing peaks often involve averages or line integrals over contiguous
sky areas, which are confounded by survey masks~\citep{0004-637X-784-1-31}.
The bias introduced by masking may be overcome with forward simulations of the nonlinear lensing
convergence field, which has been achieved with $N$-body cosmological simulations of large-scale
structure combined with ray-tracing predictions of the lensing statistics~\citep[e.g.,][]{2016ApJ...819..158B,2016arXiv160501100P}.

In a complimentary approach to cosmological parameter inference, other groups have considered a 
linear theory approximation to (suitably smoothed) lensing mass maps. 
Specifically, if (i) the shear field can be approximated as Gaussian distributed, (ii) the noise 
distribution in the galaxy ellipticity measurements (and intrinsic shape distribution) can also 
be approximated as Gaussian distributed, and (iii) the weak shear approximation is valid, then 
mass maps can be obtained using the Wiener Filter (WF)~\citep{wiener1949extrapolation}. 
The WF is a well-studied signal-inference algorithm that has been applied to cosmic microwave 
background (CMB) analysis with great success in a Bayesian context~\citep{2004PhRvD..70h3511W}.
\citet{Alsing2015} and \citet{2016arXiv160700008A} have recently extended the WF technology from
the CMB literature to the problem of weak lensing shear field and power spectrum inference. However,
these WF approaches are limited not only by the WF assumptions above but also by requirements
to smooth the measured galaxy ellipticity field to a uniform grid on the sky and to work solely
within a linear theory approximation for the distribution of cosmological mass density perturbations.
Working within the stated assumptions, the primary challenge for WF approaches is the computation
of large matrix solve operations. Novel and effective algorithms for sampling from the WF
distribution have been demonstrated for the CMB~\citep{2013A&A...549A.111E,2016ApJ...820...31R}
that are also effective for smoothed, weak-shear, and linear theory cosmic shear
inferences~\citep{2016arXiv160700008A}.


This paper is structured as follows. In \autoref{sec:method} we describe the statistical
framework for lensing convergence and shear inference given galaxy images or a galaxy
ellipticity catalog. In \autoref{sec:framework} we describe the Gaussian Process (GP)
prior for the lensing potential and how this informs the correlated inferences of shear
and convergence. We give a prescription for GP parameter optimization in
\autoref{sub:optimizing_interpolation_parameters}.
We apply our method to infer lensing convergence maps of simulated ellipticity catalogs
in \autoref{sub:simulation_study} and of an observed galaxy cluster in \autoref{sub:abell}.
We describe our main conclusions in \autoref{sec:conclusions}.
We provide details of the GP covariance derivation for lensing shear and convergence
fields in \autoref{sec:Gaussian process covariances} and provide
lensing map inferences under an analogous covariance
for a cosmological model with a linear theory approximation in
\autoref{sec:cosmology_dependent_covariance_model}.

\section{Method}
\label{sec:method}

\begin{figure*}[!htb]
  \centerline{
    \includegraphics[width=0.48\textwidth]{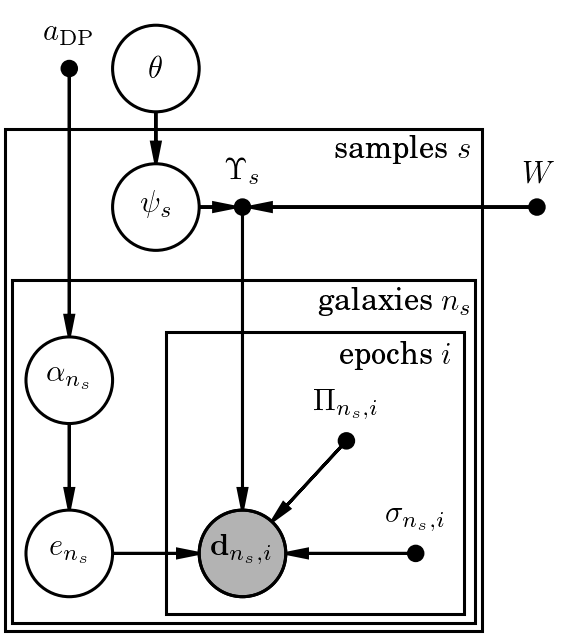}
    \includegraphics[width=0.4\textwidth]{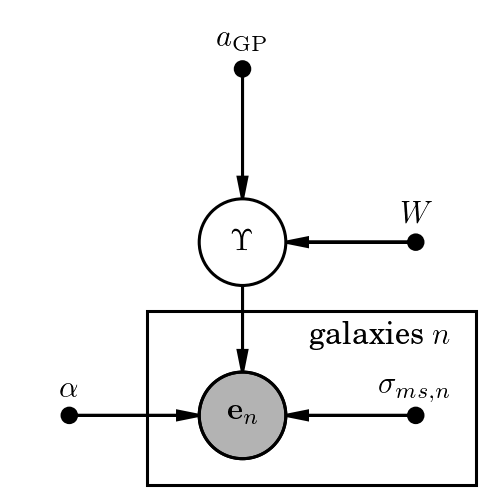}
  }
  \caption{Two successive levels of approximation for our statistical model for sampling
  probabilistic lensing shear and convergence $\lensparams$ fields.
  The unshaded circles indicate sampling parameters.
  Shaded circles indicate observed parameters while dots indicate parameters with fixed values
  rather than being sampled.
  Left panel: The model for galaxy image pixel data $\data_{n,i}$ for each galaxy $n$ and observation
  epoch $i$ requires specification of the pixel noise $\sigma_{n,i}$, the intrinsic (i.e., unlensed)
  galaxy ellipticity $e_n$, the lensing fields $\lensparams$, and the PSF at each galaxy location in
  each epoch $\psf_{n,i}$. The distribution of galaxy intrinsic ellipticities is described by the
  parameters $\alpha_n$, which can specify distinct distributions for each galaxy $n$.
  In \citet{mbi-theory} we infer marginal constraints on $\alpha_n$ under a Dirichlet Process (DP)
  prior. For this initial study, we assert a Gaussian Process (GP) prior on the lens fields
  $\lensparams$ in this paper, with the assumption that the posterior inferences of $\lensparams$
  will be related to a cosmological model in a separate analysis pipeline as we describe in the
  text. Here we assert a known PSF $\psf_{n,i}$ at every galaxy location, again deferring the
  inference and marginalization of PSFs to a separate paper. The inference of the lens fields also
  depends on our assertion of the survey window function $W$.
  Right panel: The approximate statistical model when the galaxy imaging pixel data is summarized as
  a galaxy ellipticity catalog ($e_{n}, \sigma_{e,n}$ for $n=1,\dots,\ngal$) requires specification of
  the intrinsic ellipticity distribution parameters $\alpha$ (now assumed the same for all galaxies),
  the ellipticity measurement uncertainties $\sigma_{\rm ms,n}$ per galaxy, and the lens fields
  $\lensparams$.
  We now also assert the variance, $\alpha$ of the assumed zero-mean
  intrinsic ellipticity distribution.
  }
  \label{fig:pgm}
\end{figure*}

We describe a joint probabilistic model for the lensing shear and convergence given galaxy imaging
data. We previously presented the complete statistical framework for cosmic shear
inference~\citep{mbi-theory} and here focus on the specifics of lensing convergence inference
as a function of sky coordinates in both linear and nonlinear regimes of the cosmological
mass density perturbations. In \autoref{fig:pgm} (left panel) we show the relationships in
the probabilistic model for CCD galaxy imaging data for multiple observation epochs, multiple
galaxies, and multiple galaxy samples (e.g., samples selected in different photometric redshift bins).

\begin{table*}
\begin{center}
\caption{Parameters for the statistical model. }
\label{tab:sampling_parameters}
\begin{tabular}{cl}
\hline
Parameter & Description \\
\hline
$i$ & index over epochs, or different exposures of a galaxy\\
$n$ & index over galaxies\\
$\psf_{n,i}$ & Point Spread Function (PSF) for galaxy $n_s$ in epoch $i$ \\
$e_{n}$ & Intrinsic (pre-lensing) ellipticity of a galaxy\\
$\lensparams_{s}$ & lensing shear and convergence that modify a galaxy image\\
$\lenspot_{s}$ & lensing potential \\
$\theta$ & cosmological parameters\\
$W$ & survey window function \\
$\sigma_{n,i}$ & noise properties in a galaxy image\\
$\alpha_{n}$ & parameters specifying the intrinsic ellipticity distribution for a galaxy\\
$a_{\rm DP}$ & parameters specifying the distribution over $\alpha_{n_s}$\\
${\bf d_{n}}$ & data vector (measured $e_{1,2}$ for each galaxy $n$)\\
$\sigma_{{\rm ms}; n}$ & ellipticity measurement error for galaxy $n$ \\
$\xv$ & vector of 2D spatial locations of galaxies \\
$\eint_{n}$ & intrinsic galaxy ellipticity for galaxy $n$ \\
$\alpha\equiv\sigma_{e}^2$ & parameters of the distribution of galaxy parameters \\
W & window function for the survey footprint \\
$\gpparams$ & parameters of the GP kernel\\
$\lambdagp$ & precision of the GP kernel (element of $\gpparams$) \\
$\lsqgp$ & squared GP correlation length (element of $\gpparams$) \\
\hline
\end{tabular}
\end{center}
\end{table*}

The model for observed galaxy images requires (at least) specification of the point-spread
function (PSF), $\psf_{n_s,i}$, at all $n_s$ galaxy locations in all epochs $i$, the intrinsic
or pre-lensing shape or morphology of the galaxy,
$e_{n_s}$, the applied lensing shear and magnification, $\lensparams_{s}$, and the noise properties
of the pixelated image, $\sigma_{n_s,i}$ (e.g., Gaussian with an asserted r.m.s.).

To simplify the discussion for this paper focused on shear inference, we assume the PSF $\psf_{n_s,i}$ is
known at the locations of all galaxies in all epochs. We do not further consider errors in the
PSF inference in this work. Following \citet{mbi-theory} we allow for the intrinsic ellipticity
(and potentially size, fluxes, etc.) of each galaxy to be drawn from galaxy-specific distributions
with parameters $\alpha_{n_s}$. The parameters $\alpha_{n_s}$ in turn are hierarchically 
distributed under a distribution with with parameters $a_{\rm DP}$, allowing 
inference of the effective `shape noise' in the 
shear inference. We use the label $\lensparams$ for 
the model for the lensing shear $\shear$ and convergence $\kappa$, which uniquely specify the 
components of the trace-free linear distortion matrix  under a weak lensing ($\kappa\ll1$) 
approximation of the lens equation~\citep[e.g.,][]{bartelmann01}.

As illustrated by the `plates' in the left panel of \autoref{fig:pgm}, the PSF is unique to each
observation epoch
but the intrinsic properties of a galaxy image are common across epochs (for each unique galaxy).
A key feature in the graphical model of the left panel in \autoref{fig:pgm} is that the lens fields
$\lensparams_{s}$ are common across all galaxy images in all epochs, indicating that the we require
a spatially correlated model for the coherent lensing shear and magnification patterns on the sky.
Specifying an appropriate correlated model for the lens fields is the main focus of this paper.
We allow the lens field models to be distinct for different galaxy samples $s$ in
\autoref{fig:pgm} because, e.g., galaxies in different photometric redshift bins will be lensed
by partially different foreground mass distributions. Only at the top level of the graphical
model do the lens field inferences from different galaxy samples become connected under a
cosmological model with parameters $\theta$.

We therefore explore an interim probabilistic model for the lens fields inferred from distinct
galaxy samples $s$ such that inferences of $\lensparams_s$ are statistically independent for
different $s$. This will allow us to separate computationally expensive components of a cosmic
shear inference pipeline, explore multiple cosmological and systematics models for the data,
and eventually perform rigorous uncertainty propagation and marginalization of image and intrinsic
galaxy nuisance parameters as outlined in \citet{mbi-theory}.

In measuring lensing shear of galaxy images, we must marginalize over the
intrinsic ellipticities, $\eint$, of the images. This is often done by averaging the ellipticities
of galaxies in neighboring regions of the sky and redshift, where the weights can include the
measurement and shape noise models. But, here we perform a
marginalization over an explicit intrinsic ellipticity distribution,
\begin{multline}\label{eq:marg_like}
  \prf{\data | \lensparams(\xv), \alpha} = \prod_{n=1}^{\ngal}
  \int d^{m}\eint_n\,
  \prf{\data_n | \eint_n, \lensparams(\xv_n)}
  \\ \times
  \prf{\eint_n|\alpha},
\end{multline}
where we assume the likelihood functions for each galaxy image $n=1,\dots,\ngal$ are
statistically independent~\citep[see][for more discussion of this assumption]{mbi-theory}.
The data vector $\data_{n}$ is composed of either the pixel values contributing to the image of
galaxy $n$ or a summary statistic of those pixel values.

As a pedagogical step in the development of our probabilistic lens field model,
we will consider an approximate likelihood function for summary statistics of the pixel data;
namely estimators for the
ellipticity of each galaxy image $e\equiv e_1 + i e_2$ along with an associated measurement
error per ellipticity component $\sigma_{\rm ms}$. See the right panel of \autoref{fig:pgm}.
We will develop the approximate model for ellipticity measurements as a data vector
in this paper but advocate for the more complete algorithm of \citet{mbi-theory}
for any data analysis because of the known large biases in using ellipticity estimators for
cosmic shear~\citep[e.g.,][]{Refregier++2012,Kacprzak++2012}.

Assuming Gaussian distributed ellipticity measurement errors the likelihood
function is then,
\begin{equation}\label{eq:likelihood}
  \prf{\data_n|e_n, \lensparams(\xv_n)} = \normdist_{\hat{e}_n}
  \left(e_n, \sigma^2_{{\rm ms}; n}\ident_{2}\right),
\end{equation}
where we explicitly label the data $\data_n\equiv\hat{e}_{n}$ as ellipticity estimators,
and the distribution is bivariate given the two ellipticity components.

A general likelihood function depends on the observable galaxy properties
such as ellipticity, size, and flux, which are modified by lensing from the intrinsic
properties described by $\eint$.
So, to evaluate \autoref{eq:likelihood}, we define the lensed galaxy parameters,
\begin{equation}\label{eq:lens_transform}
  e_n(\lensparams(\xv_n)) \equiv f(\eint_n, \lensparams(\xv_n)),
\end{equation}
where $f(\cdot)$ denotes the function that transforms the intrinsic, unlensed, galaxy
ellipticities $\eint_n$ to those in the lensed model $e_n$
under the action of the lensing convergence and shear specified by $\lensparams(\xv_n)$
at the galaxy sky location $\xv_n$.
In the weak shear limit defined by $\kappa \ll 1$, \autoref{eq:lens_transform} reduces to
\begin{equation}\label{eq:weak_shear}
  \tilde{e}_{n}^{\rm weak-shear} = \eint_{n} + g(\xv_n),
\end{equation}
where $g\equiv\shear/(1-\kappa)\approx\shear$ is the reduced shear.

For our pedagogy, we specify a Gaussian distribution for the \emph{unlensed}
galaxy properties $\eint$ to use in evaluating \autoref{eq:marg_like},
\begin{equation}
  \prf{\eint_{n}|\alpha,\lensparams(\xv_n)} =
  \normdist_{\eint_n} \left(0, \sigma_{e}^2\ident_{2}\right).
\end{equation}

Using the weak shear approximation in \autoref{eq:weak_shear}, we can perform
the marginalization integral in \autoref{eq:marg_like} analytically,
\begin{align}
  \prf{\data | \lensparams(\xv), \alpha} &= \prod_{n=1}^{\ngal}
    \int d^{2}\eint_n\,
    \normdist_{\eint}\left(\hat{e}_n - g(\xv_n), \sigma_{{\rm ms};n}^2\ident_2\right)
    \notag\\
    &\quad\times
    \normdist_{\eint}\left(0, \sigma_{e}^{2}\ident_{2}\right)
    \\
    &= \prod_{n=1}^{\ngal}
    \normdist_{g(\xv_n)}\left(\hat{e}_n, \left(\sigma_{{\rm ms};n}^2 + \sigma_{e}^2\right)\ident_{2}\right)
    \\
    &= \normdist_{\Matrix{g}}\left(\hat{\Matrix{e}}, \nmat\right),
    \label{eq:marg_like_analytic}
\end{align}
where in the final line we defined the $2\ngal$-length vector of observed ellipticities $\hat{\Matrix{e}}$,
the same-length vector of reduced shear components at each galaxy locations $\Matrix{g}$, and
the diagonal $2\ngal\times 2\ngal$ dimensional covariance matrix $\nmat$ with the $n$th diagonal
entry equal to $\sigma^2_{{\rm ms};n} + \sigma^2_{e}$.
Note the weak shear likelihood does not depend on the lensing convergence $\kappa$ (because the
reduced shear is approximated as equal to the non-reduced shear and we ignore lensing magnification
effects on the galaxy sizes and fluxes).

\subsection{A maximum entropy prior for lensing fields} 
\label{sec:framework}

We want an interim prior on the lensing convergence and shear that is not only
independent of cosmology but also broadly encompassing of the possible
cosmological interpretations and spatially varying systematics contributions.
That is, by choosing a functional form of a prior
for interpolating shear over the sky and marginalizing statistical uncertainties,
we should not restrict the class of physical models that might explain the data.
A mathematical version of this sentiment is that we want a
maximum entropy prior~\citep{PhysRev.106.620,shore1980} on the lensing fields.

For an assumed mean and (co-)variance, the Gaussian distribution is the maximum entropy
distribution~\citep{10.2307/2984828}. Therefore, because it is principally the second moment of the
lensing convergence that we use to constrain cosmological models, we choose a
Gaussian Process (GP) interim prior for sampling convergence (and shear) fields on
the sky given measurements of galaxy image moments.

Gravitational lensing shear is a spin-2 field and is non-locally
related to the lensing convergence, which both present modeling challenges.
However, both the lensing convergence and shear can be derived as second
derivatives of a scalar valued lensing potential $\lenspot$. So we impose the
GP prior on the potential $\lenspot$ and derive the related priors on the
convergence $\kappa$ and shear $\gamma$ from this starting assertion.

Each of the lensed observables, $\lensparams(\xv) \equiv \left[\kappa, \gamma_1, \gamma_2\right]$
is related to the lensing potential $\psi$ via derivatives in the form of:
\begin{align}
  \kappa &=\frac{1}{2}\left(\frac{\partial^2 \psi}{\partial x_1^2} +
    \frac{\partial^2 \psi}{\partial x_2^2 }\right) = \frac{1}{2}(\psi_{,11} +
    \psi_{,22})
  \label{eq:kappa_from_psi}
  \\
  \gamma_1 &=\frac{1}{2}\left(\frac{\partial^2 \psi}{\partial x_1^2} -
    \frac{\partial^2 \psi}{\partial x_2^2}\right) = \frac{1}{2}(\psi_{,11} -
    \psi_{,22})
  \label{eq:gamma1_from_psi}
  \\
  \gamma_2 &=\frac{1}{2}\left(\frac{\partial^2 \psi}{\partial x_1 \partial
    x_2} +\frac{\partial^2 \psi}{\partial x_2 \partial x_1}\right) =
    \frac{1}{2}(\psi_{,12} + \psi_{,21}).
  \label{eq:gamma2_from_psi}
\end{align}
It is straightforward to derive via integration by parts on the moments of the field 
that if $\lenspot(\xv)$ is Gaussian
distributed for given $\xv$, then so too is $\lensparams(\xv)$, but with a
modified covariance.
By specifying a GP prior on the lens potential, we therefore can derive
a GP prior on the combination of lensing convergence and shear fields that preserves the
physical correlations between these fields. We will show that we can infer the lensing 
convergence and shear via correlated draws from a GP distribution given a galaxy 
ellipticity catalog. We do not then compute lensing convergence and shear from 
spatial derivatives of a GP-distributed lens potential. The latter operation is only 
precisely defined when we have (in principle) knowledge of the lens potential at all 
sky locations. However, galaxies provide a non-contiguous background for measuring the lens 
potential. The relations in \autoref{eq:kappa_from_psi}, \autoref{eq:gamma1_from_psi}, and 
\autoref{eq:gamma2_from_psi} are then imposed only by the GP covariance structure during 
sampling and marginalization.

The derivations of using a GP to represent the lensing convergence and shear fields were
first presented in \citet{ng2016}. We refer readers to \citet{ng2016} for an introduction to
the basics of a GP. We show the derivations from \citet{ng2016} in
\autoref{sec:Gaussian process covariances} for the convenience of the reader.
In particular, the GP prior for $\lensparams$
derived from that for $\lenspot$ should not mix E and B modes in the two-point function
because we only allow for GP realizations that preserve the combinations of fields that satisfy
Equations~\ref{eq:kappa_from_psi}--\ref{eq:gamma2_from_psi} in the two-point correlations.
It is also straightforward
to extend the scalar valued potential $\lenspot$ to a complex-valued potential
$\lenspot\rightarrow\lenspot_{E} + i\lenspot_{B}$
to model or infer both E and B mode contributions to a measured shear signal.

We choose a squared exponential kernel for the GP model of the lensing potential $\lenspot$,
\begin{align}\label{eq:gp_cov}
  \smat_{\lenspot}(\xv, \yv; \lambdagp, \lsqgp) =
  \lambdagp^{-1} \exp\left(-\half \frac{s^2(\xv,\yv)}{\lsqgp}\right),
\end{align}
where $\lambdagp$ is a precision parameter that sets the amplitude of fluctuations of the GP,
$\lsqgp$ is a squared distance defining the correlation length of the GP kernel,
and the squared distance $s^2$ between pairs of galaxy locations is,
\begin{equation}\label{eq:euclid_dist}
  s^2 \equiv (\xv - \yv)^T (\xv - \yv).
\end{equation}
The squared exponential kernel is useful in interpolating smooth response functions between 
the observed galaxy locations, as we expect for low resolution or low signal-to-noise ratio (SNR)
weak lensing mass reconstructions. Exploration of other kernel choices for different mass 
reconstruction resolutions and SNRs is an interesting question that we leave for later work.
A practical motivation for our kernel choice is that the two parameters of the squared exponential 
kernel can be optimized for different data sets without particular numerical or computational 
challenges. A further justification for our smooth kernel choice comes from our intention to use the 
GP as merely an interim prior for sampling lens fields. We will later describe how these interim lens 
field realizations may be re-weighted under cosmologically informed priors.
We list the derivatives of \autoref{eq:gp_cov} in Appendix \ref{sec:Gaussian process covariances} 
that are required to build the joint covariance for the lensing shear and convergence.
We will denote the covariance constructed from second derivatives of \autoref{eq:gp_cov} as
$\smat_{\rm GP}$.

We now return to the pedagogical derivation of the likelihood function for galaxy ellipticity
estimators adding the GP distribution for the shear.
Because both the marginal likelihood in \autoref{eq:marg_like_analytic} and the shear prior are
Gaussian distributions in the shear, we can specify the shear (and convergence) joint posterior
for all galaxy and grid locations as a multivariate Gaussian distribution (otherwise known
as the Wiener filter),
\begin{multline}\label{eq:shear_posterior}
  \prf{\lensparams(\xv,\xv') | \data, \alpha, \gpparams}
  =
  \normdist_{\lensparams(\xv,\xv')}
  \left(\muv_{\lensparams}, \smat_{\lensparams} \right)
  \\ \times
  \normdist_{\hat{\Matrix{e}}}\left(0, \nmat + \smat_{\rm GP}\right),
\end{multline}
where,
\begin{align}
  \smat_{\lensparams} &\equiv \smat_{\rm GP}\left(\smat_{\rm GP} + \nmat\right)^{-1}\nmat
  \label{eq:posterior_cov}
  \\
  \muv_{\lensparams} &\equiv \smat_{\rm GP} \left(\smat_{\rm GP} + \nmat\right)^{-1} \hat{\Matrix{e}}.
  \label{eq:posterior_mean}
\end{align}
Calculation of $\muv_{\lensparams}$ in \autoref{eq:posterior_mean} yields a posterior mean
estimate of the lensing shear and convergence at every galaxy location. However, this estimator
requires inversion of an $2\ngal\times2\ngal$ covariance matrix, which can be computationally expensive.

Other works have attempted to reduce the dimensionality of the covariances
in \autoref{eq:shear_posterior} by interpolating and averaging
the measured galaxy ellipticities onto a grid of coarser resolution than that
sampled by the galaxy angular distribution~\citep[e.g.,][]{Alsing2015}.
However, such interpolations not only restrict the measured dynamic range,
but also can be expected to introduce artefacts in the inferred shear field on the grid based on
the shape of the smoothing kernel, to propagate
shape measurement systematics to a broad range of angular scales,
and to ignore error propagation from individual galaxy shape measurements to the shear inference.
Our method defines an explicit interpolation from galaxy to other sky locations, with error
propagation included.

We can marginalize the lensing fields at the galaxy locations to obtain the marginal
posterior for the lensing fields at just the smaller number of grid locations $\xv'$,
\begin{multline}\label{eq:lens_posterior}
  \prf{\lensparams(\xv') | \data, \alpha, \gpparams} \propto
  \int d\lensparams(\xv)\,
  \prf{\data | \lensparams(\xv), \alpha}
  \\ \times
  \prf{\lensparams(\xv') | \lensparams(\xv), \gpparams}
  \prf{\lensparams(\xv) | \gpparams}.
\end{multline}
When all distributions are Gaussians, \autoref{eq:lens_posterior} reduces to evaluating
\autoref{eq:shear_posterior} at only the locations of the grid $\xv'$ given the input
ellipticities $\hat{\Matrix{e}}$. Note however that we still require in \autoref{eq:posterior_mean}
the evaluation of the shear model at every galaxy location, which is then interpolated by the
WF to arbitrary sky locations. Our approach is therefore different from algorithms that require
an initial averaging of galaxy ellipticity components over local sky regions.
\autoref{eq:lens_posterior} thus specifies the interim distribution of lensing shear and convergence
that can be later propagated to cosmological model analyses.

\subsubsection{Optimizing interpolation parameters}
\label{sub:optimizing_interpolation_parameters}

We can further marginalize the lens fields $\lensparams$ at all locations $\xv$ and $\xv'$ to obtain
the marginal likelihood for the GP parameters,
\begin{equation}\label{eq:gp_marg_like}
  \prf{\data | \alpha, \gpparams} = \normdist_{\data}
  \left(0, \smat_{\rm GP} + \nmat\right).
\end{equation}
We maximize the density in \autoref{eq:gp_marg_like} to determine suitable values of the
GP parameters for interpolation of the lens fields for a given ellipticity
catalog~\citep[see a similar approach in][]{Marshall:531310}.

\autoref{eq:gp_marg_like} is informative on the GP parameters if the data vector $\data$ includes
sufficient numbers of galaxies to beat down the shape noise. In cases with smaller
signal to noise ratios we can also add a cosmologically informed prior to help constrain
the GP parameters. In this case we can replace the data vector in \autoref{eq:gp_marg_like}
with a simulation of the data vector derived from a cosmological model. Then, by marginalizing
over realizations of the simulated data vector we get a marginal prior for the GP parameters,
\begin{align}\label{eq:marg_dist_cosmo}
  \prf{\gpparams| \theta, \alpha} &=
  \int d\data^{\rm sim}\,
  \prf{\gpparams | \data^{\rm sim}, \alpha} \prf{\data^{\rm sim} | \theta},
  \notag\\
  &=
  \prf{\gpparams}
  \\
  &\times
  \int d\data^{\rm sim}\, \frac{\prf{\data^{\rm sim} | \gpparams, \alpha}}{\prf{\data^{\rm sim} | \alpha}}
  \prf{\data^{\rm sim} | \theta}.
  \notag
\end{align}
In \autoref{eq:marg_dist_cosmo} we specify a model for drawing simulated data realizations
$\data^{\rm sim}$ given cosmological parameters $\theta$. A standard approach to such a model
is to first draw a realization of the 3D mass density perturbations from a Gaussian distribution
with a cosmologically determined initial power spectrum and then to evolve the initial
mass density perturbations under a numerical (e.g., $N$-body) simulation of gravitational
evolution.

Modeling only linear density perturbation evolution, the late-time model for the
projected lensing mass density, and hence the shear, is also a Gaussian distribution,
\begin{equation}\label{eq:linear_cosmo_model}
  \prf{\data^{\rm sim}|\theta}=\normdist_{\data^{\rm sim}}(0, \Sigmamat(\theta)),
\end{equation}
for some cosmological covariance $\Sigmamat(\theta)$.
To evaluate \autoref{eq:marg_dist_cosmo}  we also must determine the evidence
\begin{equation}\label{eq:evidence}
  \prf{\data^{\rm sim}|\alpha, I} \equiv
  \int d\gpparams\, \prf{\data^{\rm sim} | \alpha, \gpparams}
  \prf{\gpparams | I},
\end{equation}
where $I$ denotes prior information on the GP parameters.
Even under a linear density perturbation approximation the evidence
cannot be calculated analytically. 

\begin{figure}[!htb]
    \centerline{
        \includegraphics[width=0.4\textwidth]{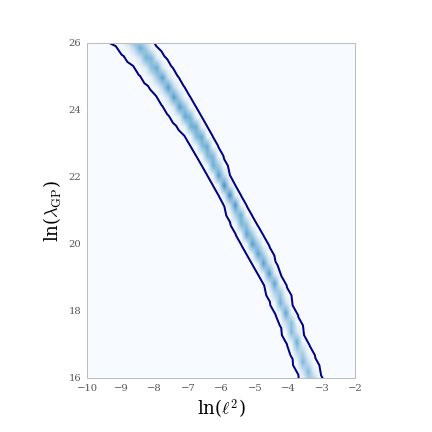}
    }
    \caption{Posterior constraints on the GP parameters after marginalizing realizations of 
    simulated lens fields under a cosmological model. The correlations and variance in the simulated 
    data serve to communicate the cosmological covariance structure into constraints on the 
    GP parameters. The tight degeneracy shows the tradeoff between low correlations and large 
    GP precision (equivalent to white noise as the GP precision becomes large) and larger correlations 
    and smaller precision as the simulated maps are fit with smoother models. The lines show the 3-$\sigma$
    confidence interval.}
    \label{fig:gp_cosmo_prior}
\end{figure}

For illustration of the GP parameter prior informed by cosmology, we use \autoref{eq:linear_cosmo_model} 
to evaluate \autoref{eq:marg_dist_cosmo} via Monte Carlo integration. We draw $N$ samples 
of $\data^{\rm sim}_{i}$ from $\prf{\data^{\rm sim}|\theta}$ and evaluate 
$\prf{\data^{\rm sim}_{i}|\gpparams, \alpha}$ for each $i=1,\dots,N$. We approximate 
the evidence in \autoref{eq:evidence} via 2D numerical integration and then calculate 
\begin{equation}
  \prf{\gpparams| \theta, \alpha} \approx \prf{\gpparams}
  \frac{1}{N} \sum_{i=1}^{N} \frac{\prf{\data^{\rm sim}_{i}|\gpparams, \alpha}}
  {\prf{\data^{\rm sim}_{i}|\alpha, I}}.
\end{equation}
We show the resulting posterior constraints on the GP parameters in \autoref{fig:gp_cosmo_prior}.
The cosmological simulation realizations impose a tight constraint in a linear combination 
of the logarithm of the two GP parameters. The constraints in \autoref{fig:gp_cosmo_prior} 
indicate the cosmological lens field simulations can be fit either with a large GP precision 
and small correlation length (i.e., essentially as white noise), or with a smaller precision 
and larger correlation lengths (i.e., as a smooth correlated map). 
We thus obtain tight constraints on the GP parameters if we can include some prior knowledge 
of the correlation length scale via $\prf{\gpparams}$.

See \autoref{sec:cosmology_dependent_covariance_model} for
a description of our linear theory cosmological model for the lensing convergence.


\subsection{Cosmological parameter inference} 
\label{sub:cosmological_parameter_inference}

Under a standard cosmological model the lensing potential $\lenspot$ is related to the
3D cosmological late-time gravitational potential $\gravpotfinal$ by a projection along
the line of sight weighted by the lensing efficiency $K(a;A_s)$,
\begin{equation}\label{eq:lenspot_proj}
  \bar{\lenspot}_{s}(\xv) \equiv
  W(\xv)
  \int dz\,
  \gravpotfinal(\xv,z)
  K(z ; Z_s),
\end{equation}
where $W(\xv)$ is the survey window function, $z$ is the cosmological redshift,
and $Z_s$ defines the redshift distribution of source galaxies for a galaxy sample $s$.
The late-time gravitational potential is in turn related to the potential of the
cosmological initial conditions $\gravpotinit$ via a deterministic function $G$
defining gravitational evolution~\citep{2013MNRAS.432..894J},
\begin{equation}\label{eq:gravpot_ic}
  \gravpotfinal(\xv, z) = G\left(\gravpotinit, \theta, z\right).
\end{equation}
We assume the initial conditions $\gravpotinit$ are Gaussian distributed with mean
zero and covariance $\covinit(\theta)$, which is entirely determined by the
potential power spectrum from inflation.

A common method to simulate $\gravpotfinal$ according to \autoref{eq:gravpot_ic} is
to run a cosmological $N$-body simulation with initial conditions $\gravpotinit$
drawn from a Gaussian distribution with a specified power spectrum.
The numerical $N$-body solver allows evaluation of the function $G$, which does not
have a known analytic form beyond low-order perturbation theory.
The standard approach to evaluate \autoref{eq:lenspot_proj} for the lens potential
in a numerical simulation is to trace light rays through the observer's light cone
given the simulated $\gravpotfinal(\xv, a)$. The bundle of light rays is
evaluated at a discrete set of sky locations to predict the lensing shear and
convergence. The predicted lens fields must then be interpolated over the sky to
galaxy locations to complete the numerical cosmological model prediction.

We have already defined a probabilistic interpolation of lens fields over the sky
via the GP. We therefore define the conditional distribution of the lens potential
at galaxy locations $\xv$ given the simulations evaluated at positions $\xv'$ via the
model of \autoref{eq:gravpot_ic} and \autoref{eq:lenspot_proj}
followed by GP interpolation,
\begin{equation}\label{eq:cond_model_comparison}
  \prf{\lenspot(\xv) | \bar{\lenspot}(\xv', \theta),\gpparams}
  = \normdist_{\lenspot} \left(
    \mu_{\lenspot}, \Sigmamat_{\lenspot}
  \right),
\end{equation}
where,
\begin{align}
  \mu_{\lenspot} &\equiv \kerngp(\xv,\xv')\kerngp^{-1}(\xv',\xv')
  \bar{\lenspot}_{s}(\xv'; \gravpotinit, \theta, A_s, W)
  \label{eq:mu_psi}
  \\
  \Sigmamat_{\lenspot} &\equiv
  \kerngp(\xv, \xv) - \kerngp(\xv,\xv')\kerngp^{-1}(\xv',\xv')\kerngp(\xv',\xv),
  \label{eq:sigma_psi}
\end{align}
define the mean and covariance for the conditional multivariate Gaussian.
With \autoref{eq:cond_model_comparison} we have thus derived a conditional probability
distribution to compare theoretical predictions of the lens potential with interim
samples of the potential drawn under the GP prior.
Said another way, our comparison of conditional posterior
samples of the lens fields with the cosmological models is mediated by the interpolation
over the sky using the GP.

By marginalizing the initial conditions realizations and the lens fields realizations
we can now derive the posterior distribution for the cosmological parameters,
\begin{multline}\label{eq:marg_post_cosmo}
  \prf{\theta | \data, \gpparams, \alpha} \propto
  \prf{\theta}
  \\ \times
  \int d\lensparams\, \int d\lensparams'\, \int d\gravpotinit\,
  \delta_{D}\left(\lensparams' - h(\gravpotinit)(\xv')\right)
  \\ \times
  \frac{\prf{\gravpotinit(\xv') | \covinit(\theta)}}
    {\prf{\lensparams' | \gpparams}}
  \\ \times
  \prf{\data | \lensparams, \alpha}
  \prf{\lensparams | \lensparams', \gpparams} \prf{\lensparams' | \gpparams},
\end{multline}
where $h(\gravpotinit)$ indicates the deterministic gravitational evolution of the
initial conditions potential to late times where the lensing is observed.

The final line of \autoref{eq:marg_post_cosmo} is the interim sampling distribution
we defined in the previous section. We can thus perform the integrals in
\autoref{eq:marg_post_cosmo} via Monte Carlo with the same interim lens field
samples that are generated in the map making algorithm.
\begin{figure}[!htb]
  \centerline{
    \includegraphics[width=0.5\textwidth]{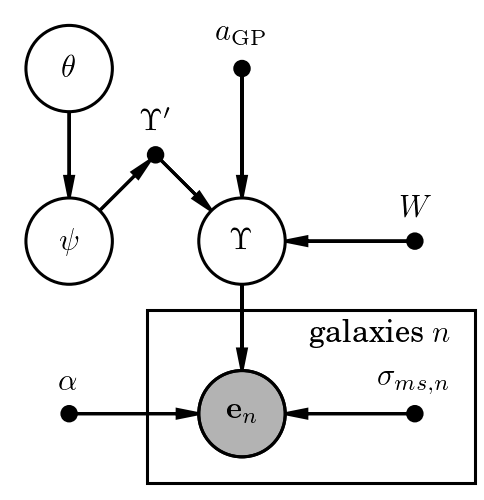}
  }
  \caption{
    In our approximate cosmological parameter inference pipeline we assume
    the lens fields at the galaxy locations $\lensparams$ are entirely determined
    by GP interpolation from the lens fields at a regular set of sky locations
    $\lensparams'$ at which we compute theory predictions.
    The lens fields on the regular set are informed by the cosmological
    model with parameters $\theta$.
  }
  \label{fig:pgm_cosmo}
\end{figure}

We propose two sampling algorithms depending on whether or not we can directly evaluate the
probability density for the lens fields under the cosmological model.
\begin{description}
  \item[Sampling algorithm 1] This is an algorithm to use when it is possible to analytically
  evaluate the cosmological probability density function.
  With $K$ samples of $\lensparams(\xv')$ from the interim sampling distribution,
  we use importance sampling to approximate the marginalizations in
  \autoref{eq:marg_post_cosmo},
  \begin{multline}\label{eq:marglikecosmo}
    \prf{\theta | \data, a_{\rm GP}, \alpha} \approx
    \prf{\theta}
    \frac{1}{K}\sum_{k=1}^{K}
    \frac{\prf{\lensparams_{k}(\xv') | \theta}}
      {\prf{\lensparams_{k}(\xv') | \gpparams}},
  \end{multline}
  where from \autoref{eq:marg_post_cosmo},
  \begin{multline}
    \prf{\lensparams(\xv') | \theta} \equiv
    \int d\gravpotinit\,
    \prf{\gravpotinit(\xv') | \covinit(\theta)}
    \\ \times
    \delta_{D}\left(\lensparams' - h(\gravpotinit(\xv'))\right).
  \end{multline}

  \item[Sampling algorithm 2] This is an algorithm to use when cosmological modeling of the lens
  fields is only possible via forward simulation.
  For most cosmological models of interest, we can only predict the model for the
  lens fields via forward simulation of the cosmological mass density perturbation evolution.
  In this case, we have no direct mechanism to evaluate the density
  $\prf{\lensparams(\xv')|\theta}$.
  Instead, we follow the steps,
  \begin{enumerate}
    \item Draw $\gravpotinit(\xv')$ from $\prf{\gravpotinit(\xv') | \covinit(\theta)}$.
    \item Compute the predicted lens fields given the drawn initial conditions
    $h(\gravpotinit)$ (e.g., via $N$-body simulation and ray-tracing prediction of the lensing
    shear and convergence).
    \item Select $K$ samples of $\lensparams(\xv)$ from the set of interim samples under the
    GP prior.
    \item Evaluate the density, $\prf{\lensparams|\lensparams', \gpparams}$, as in
    \autoref{eq:cond_model_comparison} for each sample $K$ and compute,
    \begin{equation}
      \prf{\theta | \data, \gpparams, \alpha} \approx \prf{\theta}
      \frac{1}{K} \sum_{k=1}^{K}
      \prf{\lensparams_{k}(\xv)|\lensparams(\xv',\theta), \gpparams}.
    \end{equation}
    Note that $\lensparams'$ here is that from the cosmological forward model, not that from the
    interim sampling.
  \end{enumerate}
\end{description}

For Sampling Algorithm 1, if we assume linear cosmological perturbations we can approximate,
  \begin{equation}
    \prf{\lensparams(\xv')| \theta} \approx
    \normdist_{\lensparams(\xv')}
    \left(\mathbf{0}, \Sigmamat(P_{\lensparams}(\theta))\right).
  \end{equation}
  Under this approximation we can combine the terms from \autoref{eq:marg_post_cosmo},
  \begin{equation}\label{eq:lens_field_cond_factor}
    \prf{\lensparams | \lensparams',\gpparams} \prf{\lensparams'|\theta} =
    \normdist_{\lensparams(\xv,\xv')} \left(
      0, \mmat(\xv,\xv'; \gpparams, \theta)
    \right),
  \end{equation}
  where $\mmat(\xv,\xv'; \gpparams, \theta)$ is defined in \autoref{eq:signal_cov_cosmo}.

  If we further assume, as above, that the likelihood function is a Gaussian distribution in the
  galaxy ellipticities with a linear weak shear applied, then \autoref{eq:marg_post_cosmo} becomes,
  \begin{equation}\label{eq:marg_post_cosmo_linear}
    \prf{\theta | \data, \gpparams, \alpha} \propto \prf{\theta}
    \normdist_{\data}\left(
      0, \mmat(\gpparams, \theta) + \nmat
    \right),
  \end{equation}
  with the signal covariance as defined in \autoref{eq:signal_cov_cosmo}.

  To summarize, the approximations required for \autoref{eq:marg_post_cosmo_linear} are,
  \begin{enumerate}
    \item a likelihood function that is Gaussian in the galaxy ellipticities,
    \item galaxy ellipticity measurements that are unbiased estimators of the reduced shear,
      \item weak shear ($\kappa \ll 1$),
    \item linear cosmological perturbations.
  \end{enumerate}
  We can drop assumptions 1, 2, and 3 and still use Sampling Algorithm 1. But to avoid assumption
  4 we must resort to Sampling Algorithm 2.


\section{Results}
\label{sec:results}

We evaluate the mean and covariance of the marginal posterior for the lensing shear and convergence
from \autoref{eq:shear_posterior} for a simulated data set, where we can compare with a known
truth (\autoref{sub:simulation_study}). We also analyze
the ellipticity catalog~\citep{jee2013} in the
Deep Lens Survey (DLS)\footnote{\url{http://dls.physics.ucdavis.edu}}~\citep{2002SPIE.4836...73W}
in the vicinity of a massive galaxy cluster where the
amplitudes of the lensing fields are large (\autoref{sub:abell}).

\subsection{Simulation study}
\label{sub:simulation_study}

To validate that \autoref{eq:posterior_mean} can recover the correct shear and convergence fields,
we create simulated galaxy ellipticity catalogs with artificially small shape noise so we can
measure the lensing shear and convergence to a tuneable precision.

Our procedure for simulating galaxy ellipticity catalogs is,
\begin{enumerate}
  \item Calculate a lensing convergence angular power spectrum using the cosmology theory
  code \chomp\footnote{\url{https://github.com/karenyyng/chomp}}~\citep{2013JCAP...11..009M}.
    We assume a standard
  $\Lambda$CDM cosmology with $\Omega_m=0.3$, $\sigma_8=0.8$ and a source redshift distribution
  with a narrow peak at $z=1$.
  \item Simulate Gaussian-distributed lensing shear and convergence maps on a grid using the
  code \galsim\footnote{\url{https://github.com/GalSim-developers/GalSim}}~\citep{galsim}.
  \item Place one galaxy in each grid cell of the simulated lensing shear maps. These galaxies are simply
  sources of illumination for measuring the lensing fields, not cosmologically clustered galaxies.
  \item For each galaxy, draw intrinsic ellipticity components from a 2D Gaussian distribution
  with mean 0 and a specified variance, $\sigma_e^2$.
  \item Calculate lensed ellipticities for each galaxy by adding the lensing shear to the intrinsic
  ellipticities, assuming a weak shear approximation.
  \item Save the galaxy angular sky locations and ellipticity components to a catalog file.
\end{enumerate}
Given a simulated ellipticity catalog, we find the GP parameters
$\lambdagp, \lsqgp$ that maximize \autoref{eq:gp_marg_like}.
We then evaluate \autoref{eq:posterior_mean} and \autoref{eq:posterior_cov} using the optimized
GP parameters to obtain the
marginal posterior distribution of the lensing fields $\kappa, \gamma$ at all galaxy locations
as well as on a regular grid of locations.
We expect that in practical applications to large data sets the shear will only need to be
evaluated at the galaxy locations and the converngence (or lens potential) will only need to be
evaluated at a smaller number of sky grid locations, thus reducing the overall dimensionality of
the linear system to be solved.

\begin{figure*}[!htb]
  \centerline{
    \includegraphics[width=0.54\textwidth]{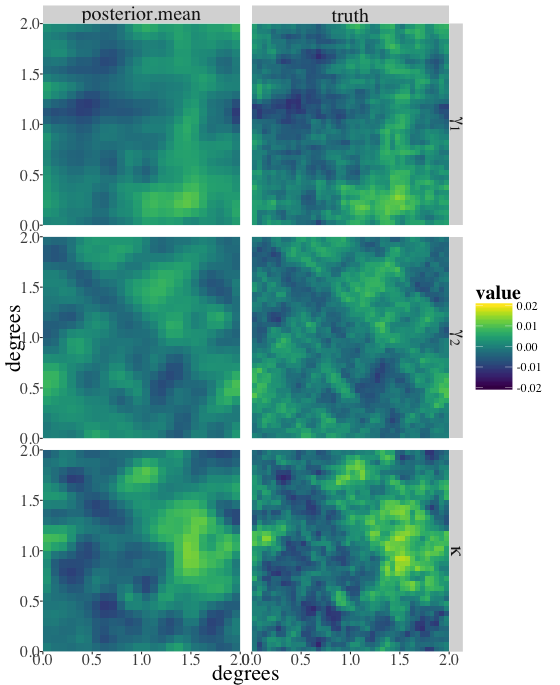}
    \includegraphics[width=0.35\textwidth]{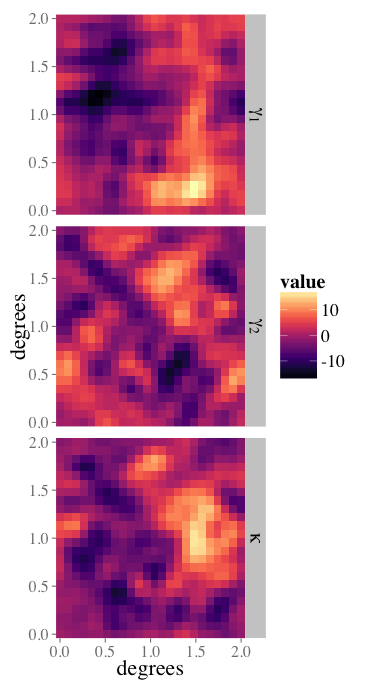}
  }
  \caption{Left: Comparison of the convergence and shear maps in our simulation study between that
  used to generate our mock galaxy ellipticity catalog (right column) and the output of our
  GP interpolation (left column). The rows show the maps for the two shear components $\gamma_{1,2}$
  and the convergence $\kappa$. These maps cover a $2\times2$ square degree field. The simulated
  intrinsic ellipticity r.m.s.\ is set to an artificially small value of $\sigma_e=0.0026$,
  which is 100 times smaller than that observed for the complete Deep Lens Survey catalog.
  Right: signal-to-noise ratio (SNR) maps for the same simulations. We calculate SNR as the
  ratio of the map to the square root of the diagonal of the covariance
  in \autoref{eq:posterior_cov}.
  These simulations use 1600 galaxies and grid sizes for the GP interpolated shear and convergence
  with 24 grid cells per dimension.
  }
  \label{fig:mass_map_comparison_galsim}
\end{figure*}

We show an example of the output of this procedure in
\autoref{fig:mass_map_comparison_galsim}
compared to the input shear fields used to generate the mock ellipticity catalog. In the
left column in \autoref{fig:mass_map_comparison_galsim} we show the posterior mean fields from
\autoref{eq:posterior_mean}.
In the adjacent panel we show the input shear fields.
We show the `true' convergence that we calculated at the same time as the input shear, but we
do not use the convergence at any point in our calculation. The ``estimated convergence'' comes
from interpolating the measured galaxy ellipticities with the GP kernel.
The right panel in \autoref{fig:mass_map_comparison_galsim} shows the
signal-to-noise ratio (SNR) of the lens field maps, defined as,
\begin{equation}\label{eq:snr_def}
  {\rm SNR} \equiv \lensparams / \sqrt{{\rm diag}\left(\smat_{\lensparams}\right)},
\end{equation}
with $\smat_{\lensparams}$ defined in \autoref{eq:posterior_cov}.
In the limit that the variance of the lens fields dominates the intrinsic shape
and ellipticity measurement variances, \autoref{eq:posterior_cov} reduces to
$\nmat$. Then \autoref{eq:snr_def} becomes
${\rm SNR} \rightarrow \lensparams / \sqrt{\sigma_e^2 + \sigma_{\rm ms}^{2}}$,
which is similar to SNR definitions in other weak lensing mass mapping
analyses~\citep{2000MNRAS.313..524V,2014MNRAS.442.2534S}.

The simulation in
\autoref{fig:mass_map_comparison_galsim} has an artificially low intrinsic ellipticity r.m.s.\
of $\sigma_e=0.0026$ compared to $\sigma_e=0.26$ in the Deep Lens Survey that
we analyze in \autoref{sub:abell}.
We choose a small shape noise r.m.s.\ to
allow us to validate our convergence inference with a small number of only 1600 galaxies.
We place the simulated galaxies on a $40\times40$ grid, so that we do not have to interpolate
the simulated shear fields to build the mock ellipticity catalog.

Because the Gaussian model for the shape noise r.m.s.\ scales with the number of galaxies
$\ngal$ as $\ngal^{-1/2}$, the shape noise r.m.s.\ in
our simulation is equivalent to a DLS-like galaxy sample with 16 million galaxies.
This is about 64 times the number of galaxies we have in a single four square degree field
of the DLS. We show the effect of increasing the shape noise r.m.s.\ by a factor
of $\sqrt{64}$ later in this section.

The reconstructed shear and convergence maps in \autoref{fig:mass_map_comparison_galsim} are a
good match to the simulation inputs, but are noticeably smoothed. This is for two reasons;
(1) we evaluate the interpolated lens fields on a smaller $24\times24$ grid than the
$40\times40$ grid on which the inputs are evaluated,
(2) we use a value of $\lsqgp=0.0123$ that is larger than the Nyquist scale in the maps, even
for the $24\times24$ interpolated grid because of the way the GP parameters are optimized.
We will discuss the GP parameter optimization below.

\begin{figure}[!htb]
  \centerline{
    \includegraphics[width=0.5\textwidth]{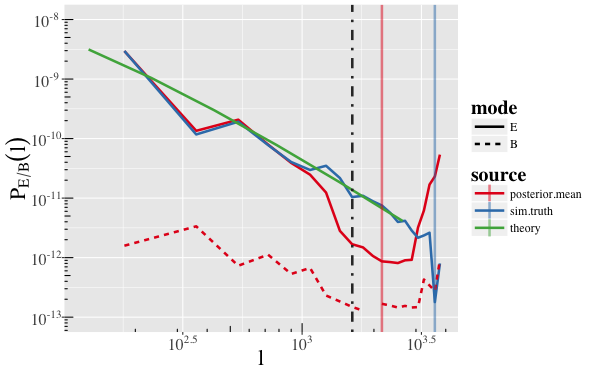}
  }
  \caption{Comparison of power spectrum estimators in our simulation study to the `truth' input
  spectrum.
  The vertical solid lines show the Nyquist frequencies of the grids.
  The vertical dot-dashed line shows the multipole corresponding to the length scale
  $\sqrt{\ell_{\rm GP}^{2}}$ set by the GP kernel parameter $\ell_{\rm GP}^{2}=0.0123$ for
  field coordinates are normalized to the unit square.
  }
  \label{fig:power_spectra_galsim}
\end{figure}

In \autoref{fig:power_spectra_galsim} we show the E and B mode power spectrum estimators obtained
from the posterior mean shear fields shown in \autoref{fig:mass_map_comparison_galsim}.
We compare the E-mode power spectrum estimator from the mean posterior shear maps with that
using the higher-resolution shear maps that were used to generate the mock ellipticity
catalog (which we label `simulation truth'). We also show in \autoref{fig:power_spectra_galsim}
the `theory' power spectrum that we used to generate the `simulation truth' shear maps.
The `theory' and `simulation truth' spectra agree on scales below the Nyquist frequency,
shown in \autoref{fig:power_spectra_galsim} as the vertical blue line. The mean posterior
power spectrum agrees with the `simulation truth' spectrum on scales below the effective
smoothing frequency derived from the value of $\lsqgp$ and shown by the vertical
dot-dashed line in \autoref{fig:power_spectra_galsim}.

\begin{figure*}
  \centerline{
    \includegraphics[width=0.6\textwidth]{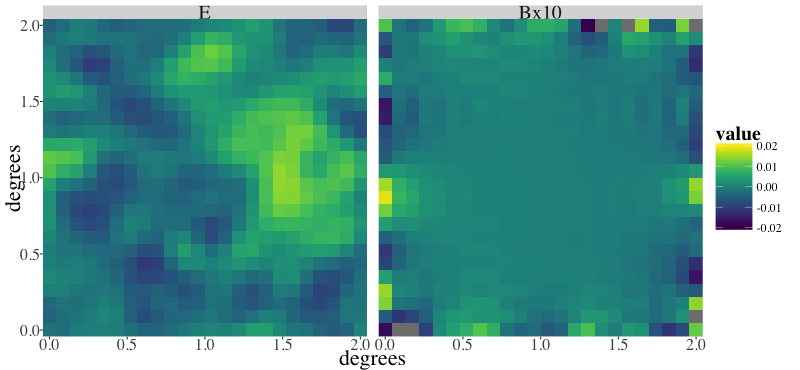}
  }
  \caption{E and B mode maps derived from the posterior mean shear maps shown in the left column
  of \autoref{fig:mass_map_comparison_galsim}.
  The E-mode map closely matches the interpolated convergence map in
  \autoref{fig:mass_map_comparison_galsim} as expected.
  The B-mode map is near zero throughout the center of the field, but shows non-zero
  values around the field edge because of mathematical ambiguities in the E/B mode separation
  at the field boundaries~\citep{bunn2003}. Note the value of the plotted B-mode map
  has been multiplied by 10 for easier visualization.
  }
  \label{fig:EB_maps}
\end{figure*}

The B-mode power spectrum estimated
from the mean posterior shear maps is shown by the dashed red line in \autoref{fig:power_spectra_galsim}.
We expect the B-mode power to be consistent with zero because we simulated only E-mode power.
The nonzero B-mode power spectrum in \autoref{fig:power_spectra_galsim} is explained by
examining \autoref{fig:EB_maps}, which shows the E and B mode mean posterior maps from which the
E and B mode mean posterior power spectra were derived. The B-mode map in \autoref{fig:EB_maps}
is near zero throughout the field except near the boundaries. These edge effects in the B-mode map
can be explained by mathematical ambiguities in the definition of E and B mode separation in a
finite field~\citep{bunn2003}. The small value of the B-mode map in \autoref{fig:EB_maps} away from
the field boundaries indicates our GP interpolation method does not create spurious B-modes.

\begin{figure*}
  \centerline{
    \subfigimg[width=0.3\textwidth,pos='LOWER LEFT',hsep=15pt,font=\footnotesize]{\textcolor{white}{$\sigma_{e}=0.0026$}}{kappa_galsim_sigmae1.png}
    \includegraphics[width=0.3\textwidth]{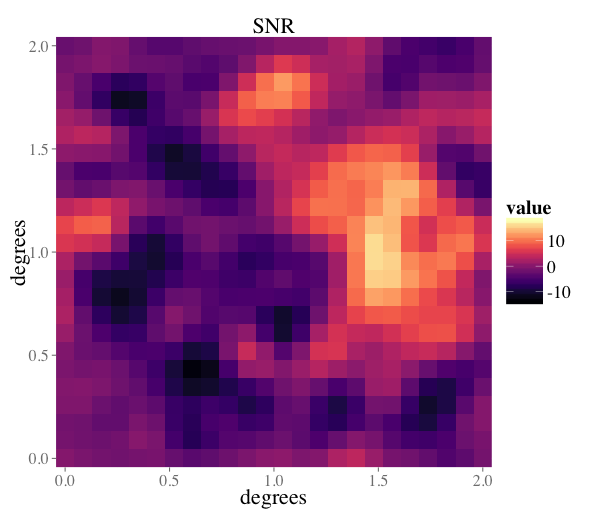}
    \subfigimg[width=0.3\textwidth,pos='LOWER LEFT',hsep=15pt,font=\footnotesize]{\textcolor{white}{$\lambda_{\rm GP}=2.2\times 10^{8}, \ell^2=0.012$}}{lnp_galsim_sigmae1.png}
  }
  \centerline{
    \subfigimg[width=0.3\textwidth,pos='LOWER LEFT',hsep=15pt,font=\footnotesize]{\textcolor{white}{$\sigma_{e}=0.0077$}}{kappa_galsim_sigmae2.png}
    \includegraphics[width=0.3\textwidth]{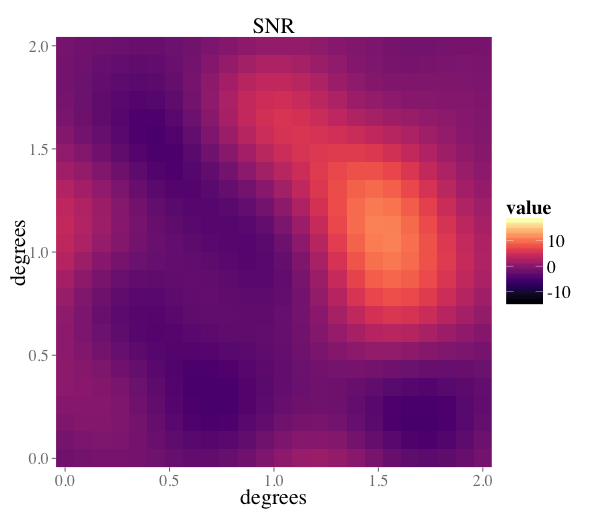}
    \subfigimg[width=0.3\textwidth,pos='LOWER LEFT',hsep=15pt,font=\footnotesize]{\textcolor{white}{$\lambda_{\rm GP}=2.3\times 10^{7}, \ell^2=0.061$}}{lnp_galsim_sigmae2.png}
  }
  \centerline{
    \subfigimg[width=0.3\textwidth,pos='LOWER LEFT',hsep=15pt,font=\footnotesize]{\textcolor{white}{$\sigma_{e}=0.013$}}{kappa_galsim_sigmae3.png}
    \includegraphics[width=0.3\textwidth]{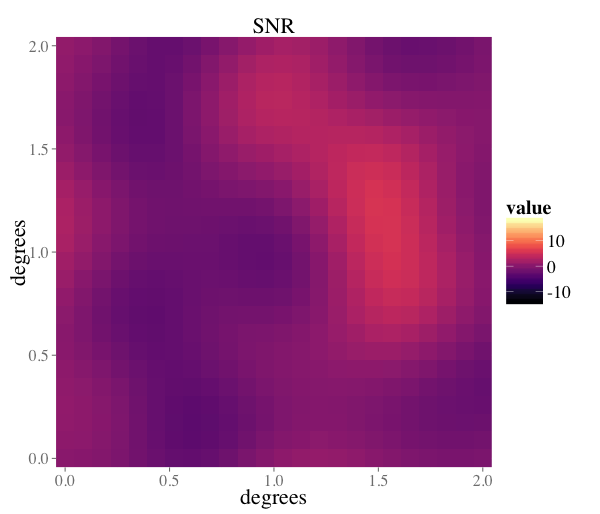}
    \subfigimg[width=0.3\textwidth,pos='LOWER LEFT',hsep=15pt,font=\footnotesize]{\textcolor{white}{$\lambda_{\rm GP}=2.3\times 10^{7}, \ell^2=0.061$}}{lnp_galsim_sigmae3.png}
  }
  \centerline{
    \subfigimg[width=0.3\textwidth,pos='LOWER LEFT',hsep=15pt,font=\footnotesize]{\textcolor{white}{$\sigma_{e}=0.023$}}{kappa_galsim_sigmae5.png}
    \includegraphics[width=0.3\textwidth]{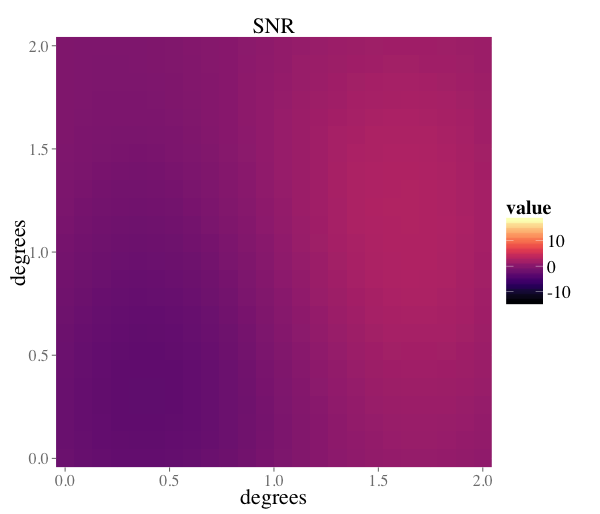}
    \subfigimg[width=0.3\textwidth,pos='LOWER LEFT',hsep=15pt,font=\footnotesize]{\textcolor{white}{$\lambda_{\rm GP}=2.3\times 10^{6}, \ell^2=0.30$}}{lnp_galsim_sigmae5.png}
  }
  \caption{Posterior mean convergence maps (left), signal-to-noise ratio (SNR) maps (middle),
  and marginal likelihood contours (right) for simulated galaxy catalogs with varying intrinsic
  ellipticity r.m.s.\ $\sigma_e$.
  From top to bottom, $\sigma_e = \left(0.00258, 0.00774, 0.0129, 0.0232\right)$ (these values are
  also annotated in the bottom left corner of each convergence map).
  The top row matches the simulation shown in \autoref{fig:mass_map_comparison_galsim}.
  Following the right column from top to bottom shows how the maximum-likelihood estimates for the
  GP parameters shifts to longer correlation lengths and smaller precision parameters with increasing
  shape noise. That is, as the data becomes more noise dominated, the marginal GP parameter likelihood
  changes shape to prefer smoothing, or effectively averaging, more galaxies to retain a more
  significant shear and convergence signal. The ML values for the GP parameters are listed
  in each panel showing the log-likelihood contours.
  }
  \label{fig:sigma_e_maps}
\end{figure*}

To illustrate the GP parameter optimization procedure we show in \autoref{fig:sigma_e_maps} the
mean posterior convergence, the associated SNR maps, and the marginal likelihood surface for
the GP parameters (left to right columns) for increasing intrinsic ellipticity variance
$\sigma_e^2$ (top to bottom rows). The top row of panels in \autoref{fig:sigma_e_maps} show a repeat
of the $\sigma_e^2$ and $\gpparams$ values used in \autoref{fig:mass_map_comparison_galsim}.
For the very small shape noise in this case, there is a narrow peak in the log-likelihood surface
for the two GP kernel parameters. We thus select the maximum likelihood (ML) values for $\gpparams$
and obtain the convergence and SNR maps that closely resemble the `true' convergence
as shown in \autoref{fig:mass_map_comparison_galsim}. However, the ML value of $\lsqgp$ is somewhat
larger than the Nyquist frequency of the grid to which we interpolate as shown in
\autoref{fig:power_spectra_galsim} (compare dot-dashed to red solid vertical lines). Because the
noise is sub-dominant in this example, we would expect a value of $\lsqgp$ matching the grid
Nyquist frequency to yield a more accurate convergence map reconstruction. We are likely to
obtain more accurate results, therefore, if we impose a prior on $\gpparams$ that encodes this
expectation.

As the shape noise increases (for the same input signal and measurement uncertainties), the peak
in the marginal log-likelihood surface for $\gpparams$ becomes broader and eventually disappears
as shown by the rows of panels from top to bottom in \autoref{fig:sigma_e_maps}. The ML value
for $\lsqgp$, while less well defined, continues to yield maps that are more smoothed as
$\sigma_e^2$ increases. This helps to preserve large amplitudes of the peaks in the SNR maps (see
the color bar scales in the SNR maps of \autoref{fig:sigma_e_maps}), but
in compensation erases structures at all but the lowest spatial frequencies in the maps. This
procedure, with flat priors in the log of $\gpparams$, appears useful for visualizing the posterior
convergence maps, but is undesirable for subsequent cosmological analyses. We see again that
we would prefer a prior favoring smaller $\lsqgp$ even as the shape noise becomes large.

\begin{figure*}
  \centerline{
    \subfigimg[width=0.3\textwidth,pos='LOWER LEFT',hsep=15pt,font=\footnotesize]{\textcolor{white}{$\sigma_{e}=0.0026$}}{kappa_galsim_prior_sigmae1.png}
    \includegraphics[width=0.3\textwidth]{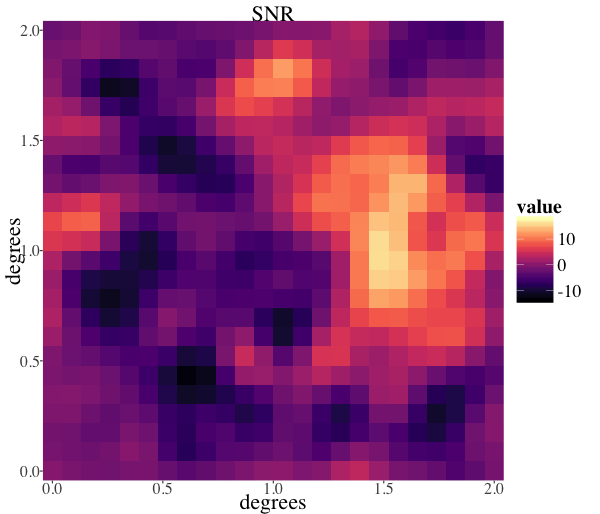}
    \subfigimg[width=0.3\textwidth,pos='LOWER LEFT',hsep=15pt,font=\footnotesize]{\textcolor{white}{$\lambda_{\rm GP}=2.2\times 10^{8}, \ell^2=0.012$}}{lnp_galsim_prior_sigmae1.png}
  }
  \centerline{
    \subfigimg[width=0.3\textwidth,pos='LOWER LEFT',hsep=15pt,font=\footnotesize]{\textcolor{white}{$\sigma_{e}=0.0077$}}{kappa_galsim_prior_sigmae2.png}
    \includegraphics[width=0.3\textwidth]{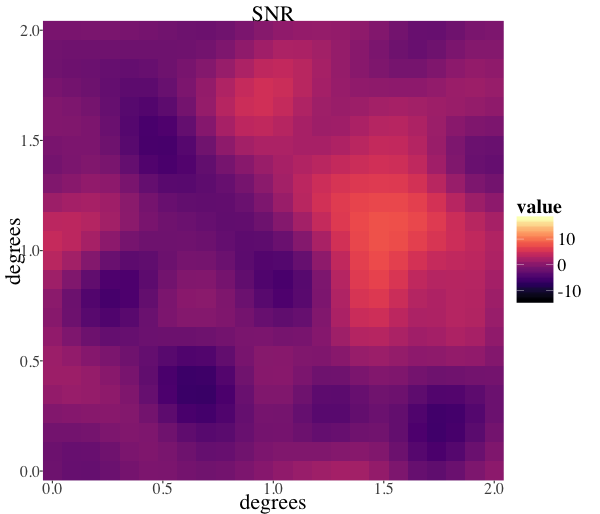}
    \subfigimg[width=0.3\textwidth,pos='LOWER LEFT',hsep=15pt,font=\footnotesize]{\textcolor{white}{$\lambda_{\rm GP}=2.3\times 10^{7}, \ell^2=0.027$}}{lnp_galsim_prior_sigmae2.png}
  }
  \centerline{
    \subfigimg[width=0.3\textwidth,pos='LOWER LEFT',hsep=15pt,font=\footnotesize]{\textcolor{white}{$\sigma_{e}=0.013$}}{kappa_galsim_prior_sigmae3.png}
    \includegraphics[width=0.3\textwidth]{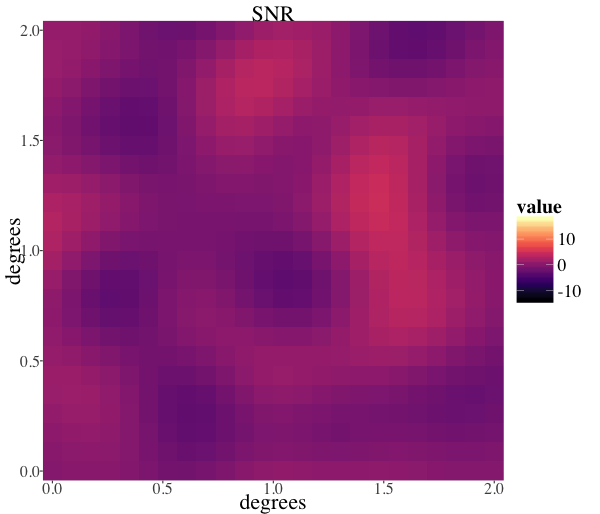}
    \subfigimg[width=0.3\textwidth,pos='LOWER LEFT',hsep=15pt,font=\footnotesize]{\textcolor{white}{$\lambda_{\rm GP}=2.2\times 10^{8}, \ell^2=0.027$}}{lnp_galsim_prior_sigmae3.png}
  }
  \centerline{
    \subfigimg[width=0.3\textwidth,pos='LOWER LEFT',hsep=15pt,font=\footnotesize]{\textcolor{white}{$\sigma_{e}=0.023$}}{kappa_galsim_prior_sigmae5.png}
    \includegraphics[width=0.3\textwidth]{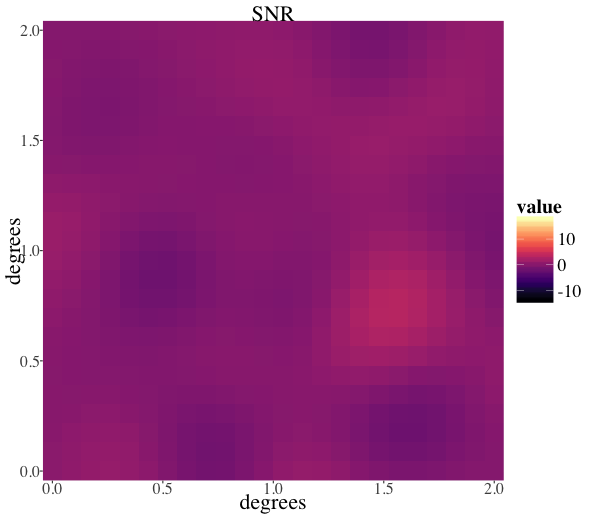}
    \subfigimg[width=0.3\textwidth,pos='LOWER LEFT',hsep=15pt,font=\footnotesize]{\textcolor{white}{$\lambda_{\rm GP}=2.2\times 10^{8}, \ell^2=0.027$}}{lnp_galsim_prior_sigmae5.png}
  }
  \caption{Same as \autoref{fig:sigma_e_maps} except with a prior asserted for the GP
  parameters $\gpparams$ that limits the degree of smoothing in the posterior mean maps.
  }
  \label{fig:sigma_e_maps_with_prior}
\end{figure*}

We assert such a prior in \autoref{fig:sigma_e_maps_with_prior}, by imposing Gaussian priors
separately in $\ln(\lambdagp)$ and $\ln(\lsqgp)$ with parameters given in
\autoref{tab:prior_params_galsim}. Our Gaussian prior is informed by the cosmological simulation study 
shown in \autoref{fig:gp_cosmo_prior} combined with our prior that the GP correlation length be large enough 
so that white noise does not dominate the fits to the lens fields. 
For each value of $\sigma_e$ in \autoref{fig:sigma_e_maps_with_prior}
we see that the signal-to-noise ratio (SNR) is comparable to that in \autoref{fig:sigma_e_maps}
but the convergence maps include higher spatial frequency structures.
\begin{table}[!htb]
  \begin{center}
    \caption{\label{tab:prior_params_galsim}Parameters for the Gaussian prior on the
      GP parameters for our simulation study.}
    \begin{tabular}{lcc}
      \hline\hline
      Parameter & mean & std. dev. \\
      \hline
      $\ln\left(\lambdagp\right)$ & $18$ & 0.43 \\
      $\ln\left(\lsqgp\right)$    & $-4.0$ & 0.1\\
      \hline\hline
    \end{tabular}
  \end{center}
\end{table}

\subsubsection{Cosmological parameter constraints}
\label{sub:cosmo_params}

Although we assert a non-cosmological GP prior on the lens fields to infer regular convergence
and shear maps, we now demonstrate how we can recover cosmological information in a manner
similar to common algorithms in the literature. That is, we compute an angular power spectrum
estimator for the lensing convergence from the lens field posterior distribution.

The posterior distribution for the lens fields given the GP parameters is a multivariate
Gaussian distribution characterized by a mean field and covariance as given in
\autoref{eq:posterior_mean} and \autoref{eq:posterior_cov}. The posterior mean is thus a
convenient and useful summary statistic (as we have shown above).
Also, because an isotropic Gaussian random field is fully described by the angular power spectrum,
it is common in the literature to reduce cosmological large-scale structure statistics
to two-point function estimators for cosmological parameter estimation.

We showed the angle-averaged (E-mode) convergence power spectrum estimator for our low-noise
Gaussian simulated maps in \autoref{fig:power_spectra_galsim}. We use this power spectrum
estimator as a summary statistic derived from the observed ellipticity catalog. We further
assert a multivariate Gaussian likelihood function for the power spectrum estimator
with covariance~\citep[e.g.,][]{2001ApJ...554...56C},
\begin{equation}
  \Cov\left(P_{\kappa}(\ell)\right) = \frac{2}{N_{\ell}}P_{\kappa}^{2}(\ell),
\end{equation}
where $N_{\ell}$ is the number of modes contributing to the band power estimator for a multipole
bin centered at $\ell$.

\begin{figure}
  \centerline{
    \includegraphics[width=0.47\textwidth]{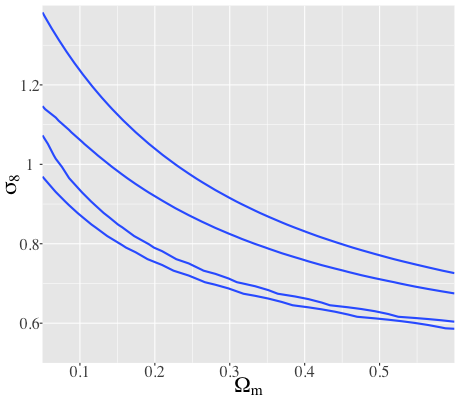}
  }
  \caption{Marginal constraints on the cosmological parameters using the E-mode power spectrum
  estimator derived from the mean posterior convergence field.}
  \label{fig:cosmo_param_contours}
\end{figure}

In \autoref{fig:cosmo_param_contours} we show 68\% and 95\% contours of the 2D posterior
distribution for the cosmological mass density $\Omega_m$ and density fluctuations r.m.s.\
$\sigma_8$ given the angular power spectrum likelihood just described and flat priors.
The values used to generate the mock data are $\Omega_m=0.3, \sigma_8=0.8$.
We limit the multipole range of the power spectrum in the likelihood to $100 < \ell < 1300$,
where the upper bound is set by the effective smoothing length imposed by the asserted GP correlation
parameter $\lsqgp$.

We see from \autoref{fig:cosmo_param_contours} that the GP interim prior used to derive the
convergence map from the galaxy ellipticity catalog has not biased the cosmological
parameter constraints (within the uncertainties) obtained from a reduced summary statistic
of the lens field posterior. We defer to later work more complete demonstrations of the
cosmological parameter inference algorithms in \autoref{sub:cosmological_parameter_inference}
based on marginalizing the lens field realizations.

\subsection{Abell cluster in the Deep Lens Survey}
\label{sub:abell}

We apply our lens field inference algorithm to the galaxy ellipticity catalog derived from the
DLS\footnote{\url{http://dls.physics.ucdavis.edu}}~\citep{2002SPIE.4836...73W}.
The DLS is a 20 square degree optical imaging survey optimized for cosmic shear measurements.
We analyze a $\sim1$~square~degree field centered on the Abell cluster 781, which was
previously analyzed using DLS lensing shear measurements in
\citet{2006ApJ...643..128W,2009ApJ...702..980K,wittman2014}.
Abell 781 consists of four massive galaxy clusters, which is a useful case study for our
algorithm because it provides a large lensing signal with a modest number of galaxies
and because the distribution of mass density perturbations is decidedly not Gaussian
distributed.

\citet{jee2013} presented galaxy ellipticity measurements with the DLS $R$-band imaging calibrated
to produce shear estimates with biases well below the statistical uncertainties for two-point
cosmic shear correlation function estimators. The DLS shear pipeline in \citet{jee2013} includes
correlated PSF size and ellipticity corrections in each DLS exposure, calibration of additive
and multiplicative shear biases with image simulations, and a set of null tests validating the PSF
shear calibration corrections. The shear estimation pipeline used in \citet{jee2013}, \textsf{sFIT},
was further validated as the winning algorithm in the blinded community shear measurement
challenge \textsf{GREAT3}~\citep{great3-paper1}.
The DLS was performed with four optical pass-bands ($BVRz$) that allow photometric
redshift (photo-$z$) estimates for all lensing source galaxies~\citep{schmidt2013}.
\citet{jee2015} extended the DLS shear analysis to a tomographic cosmic shear measurement using
the photo-$z$ estimates.

The DLS galaxy ellipticity catalog produced for \citet{jee2013} includes a catalog-level selection
based on measured galaxy magnitudes, sizes, ellipticity measurement error, photo-$z$ estimates, and
proximity to masks as listed in \citet{jee2013} Table 2.
We perform a further set of selections on this catalog as listed in
\autoref{tab:dls_selection}.
\begin{table}[!htb]
  \begin{center}
    \caption{\label{tab:dls_selection}Selection criteria applied to the DLS galaxy catalog.}
    \begin{tabular}{ccc}
    \hline\hline
    Parameter        & min         & max \\
    \hline
    $z_b$            & 0.45        & --  \\
    $de$             & --          & 0.1 \\
    $\sqrt{a^2+b^2}$ & $0.8^{''}$  & --  \\
    $R$              & 22          & 23  \\
    \hline\hline
    \end{tabular}
  \end{center}
\end{table}
We choose the lower bound on (maximum posterior) photo-$z$, $z_b$, to select source galaxies
that are likely to be at redshifts larger than that of the highest redshift sub-cluster in the field
at $z\approx0.43$~\citep{wittman2014}. We further select only those galaxies in the ellipticity
catalog with ellipticity measurement errors, $de$, less than 0.1 and sizes greater than 0.8~arcseconds to
obtain galaxies likely to have more precisely measured shapes for informing the lensing shear.
We exclude galaxies with sizes, as determined from the geometric mean of the semi-major and semi-minor axes $a,b$,
less than 0.8 arcseconds because the ellipticities tend to be less
well measured when the galaxy size is similar to that of the PSF.
We select the brighter galaxies based on $R$-band magnitude that are still likely to be faint
enough to avoid significant contamination from cluster members. After all the selections listed
in \autoref{tab:dls_selection} we measure an ellipticity r.m.s.\ of $\sigma_e=0.21$. This measurement
includes the ellipticities with lensing effects included, but because lensing is sub-dominant
to the intrinsic ellipticity dispersion we assert $\sqrt{\alpha} = \sigma_e=0.21$ for the posterior
inference for A781.

\begin{figure*}[!htb]
	\centerline{
		\includegraphics[width=0.47\textwidth]{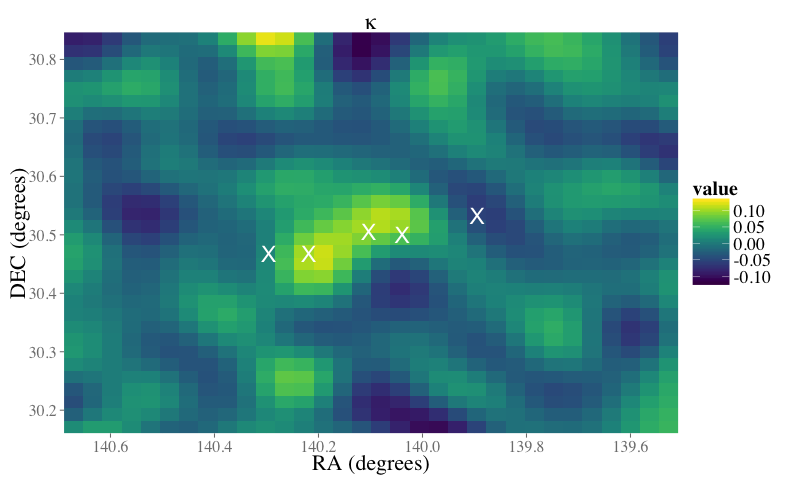}
    \includegraphics[width=0.47\textwidth]{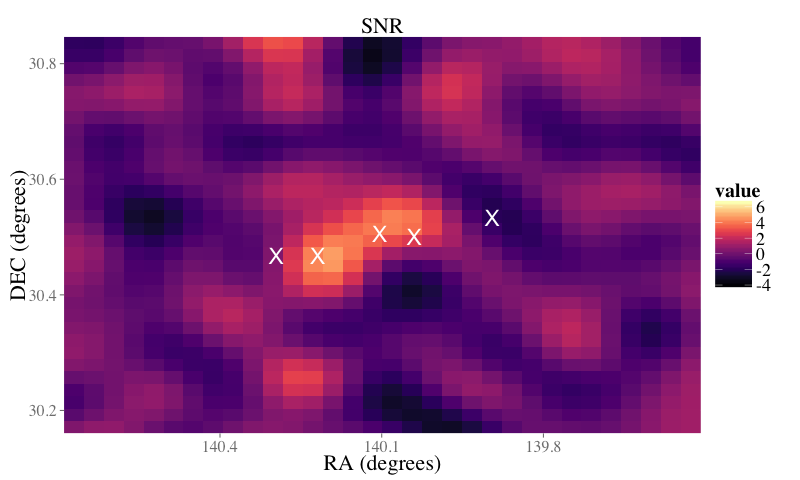}
	}
  \centerline{
    \includegraphics[width=0.47\textwidth]{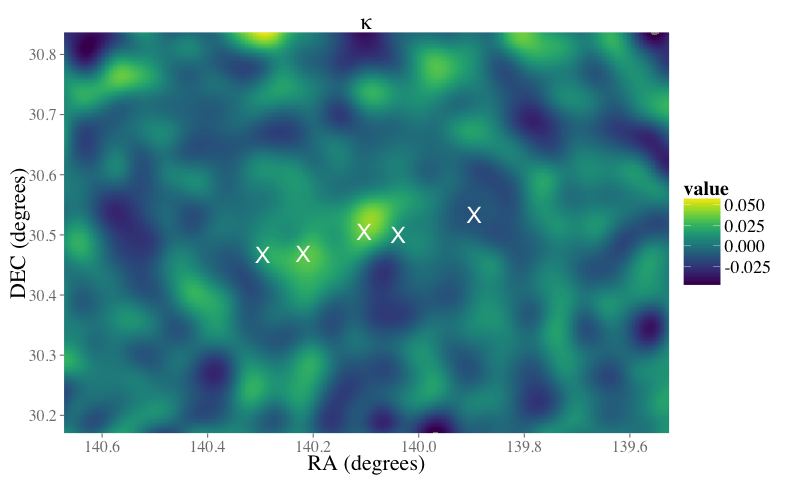}
    \includegraphics[width=0.47\textwidth]{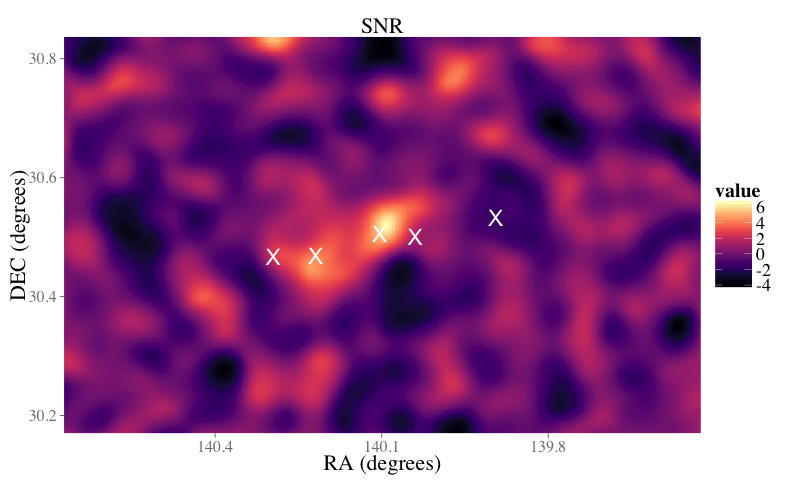}
  }

	\caption{Posterior mean lens field maps (left) and SNR maps (right) for a
  field centered on the galaxy cluster Abell 781. We use 6000 galaxies selected
  to have photometric redshifts larger than the known redshifts of the two primary
  clusters in this field of view ($z=0.296$ for Abell 781 and $z=0.43$ for a chance
  alignment of a second cluster).
  The top row of panels show the posterior mean map and SNR using our GP
  prior. The bottom row of panels show the algorithm of \citet{2012ApJ...747L..42D}
  applied to the same galaxies with the same uniform per-galaxy weighting
  as in the top row, but evaluated on a finer grid. The normalization of the
  convergence in the lower left panel is arbitrary~\citep[see][]{2012ApJ...747L..42D}.
  The white crosses indicate sub-clusters identified by~\citet{sehgal2008} with
  x-ray detections. There is no associated mass for the white cross second
  from the right in any published analyses of this system.
	}
	\label{fig:mass_map_Abell}
\end{figure*}

We show our mean posterior inference of the lensing convergence of A781 in
\autoref{fig:mass_map_Abell} using 6000 galaxies randomly selected from the cut sample described
in \autoref{tab:dls_selection}. We select only 6000 galaxies to limit the size of the lens field
joint covariance that we must invert.
The left column of panels shows the convergence while the right
column shows the signal-to-noise ratio for the mean posterior. Our calculation is in the
top row of panels in \autoref{fig:mass_map_Abell}, which we compare with the aperture mass
algorithm of \citet{2012ApJ...747L..42D} in the bottom row of panels.
Note we use the same galaxy sample as input to each mass mapping algorithm.
The algorithm of \citet{2012ApJ...747L..42D},
called `aperture densomitry', provides a mass estimator that is more localized on the sky. The shear
in apertures is averaged with weighting functions that account for both angular selections and
the expected line-of-sight lensing kernel with the aide of the photometric redshift information
in the DLS catalog. However, to make a more direct comparison with the algorithm in this paper,
we recomputed the aperture densomitry weights without using any photometric redshift information
for the source galaxies.

\autoref{fig:mass_map_Abell} shows that we obtain consistent results for the two main A781
sub-clusters using our mean posterior map and the method used in a previous analysis. The white
crosses in \autoref{fig:mass_map_Abell} indicate the locations of all sub-clusters detected
in \citet{wittman2014}. We do not detect all the same sub-clusters, which is likely
because we use a significantly
smaller number of galaxies (6000 versus $\sim 50000$) while we test the performance and scaling
of our codes. However one sub-cluster denotedin \autoref{fig:mass_map_Abell} is only detected
in the literature in x-ray emission~\citep{sehgal2008}.
\citet{wittman2014} also weight the source galaxies according to the expected
lensing kernel and the photo-$z$ estimates. We make no use of photo-$z$ information other than
in the sample selection.

\begin{figure}[!htb]
    \centerline{
        \includegraphics[width=0.45\textwidth]{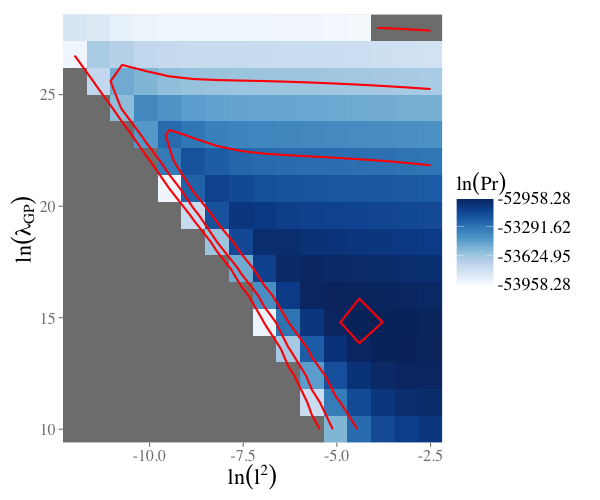}
    }
    \caption{The marginal posterior distribution for the GP parameters with
    priors imposed to favor correlation lengths smaller than the expected cluster sizes.}
    \label{fig:abell_lnp}
\end{figure}

\begin{table}[!htb]
  \begin{center}
    \caption{\label{tab:prior_params_abell}Parameters for the Gaussian prior on the
      GP parameters for A781.}
    \begin{tabular}{lcc}
      \hline\hline
      Parameter & mean & std. dev. \\
      \hline
      $\ln\left(\lambda_{\rm GP}\right)$ & $\ln\left(10^{6}\right)$ & 0.1 \\
      $\ln\left(\ell^2\right)$           & $\ln\left(10^{-4}\right)$ & 0.5\\
      \hline\hline
    \end{tabular}
  \end{center}
\end{table}
In \autoref{fig:abell_lnp} we show the marginal log-likelihood for the A781 ellipticities in
the plane of the two GP parameters. Unlike in \autoref{fig:sigma_e_maps} the noise covariance
is now significant in defining the contours in \autoref{fig:abell_lnp} such that a large
GP precision and small GP correlation length is favored (indicating a sub-dominant
signal covariance). We therefore impose a Gaussian prior in the logarithm of the GP parameters
with parameters listed in \autoref{tab:prior_params_abell}. We infer maximum posterior values
of $\lambda_{\rm GP}=2.7\times10^{6}$ and $\ell^2=0.012$, which we use in the convergence inference
in \autoref{fig:mass_map_Abell}. However, as shown in \autoref{fig:abell_lnp}, the marginal
posterior for the GP parameters is only weakly peaked for this data set, and a range of GP
parameters would be acceptable for the mass map inference.

\section{Conclusions} 
\label{sec:conclusions}

We have demonstrated a probabilistic model inference of lensing shear and convergence
using a Gaussian Process (GP) prior. Our method is an extension of previous Bayesian maximum entropy
mass mapping methods that is applicable in both cosmic shear (i.e., `field') and cluster
lensing regimes. We validated our algorithm using simulated Gaussian distributed
shear maps with comparisons to the input `truth' and by reconstructing the mass distribution
in a cluster field in the Deep Lens Survey where we compare with the results of a previously
published `aperture densometry' method.

Because our GP kernel for the lensing shear and convergence is derived from a consistent lens
potential, we find excellent separation of E and B modes in the posterior lens field maps. 
Recently \citet{2016arXiv161003345B} presented an algorithm to project Wiener Filter maps into 
`pure' E and B modes, removing ambiguity in the E/B decomposition from survey masks. Because we compute 
Wiener Filter solutions for the lensing convergence from galaxy ellipticities, the method of 
\citet{2016arXiv161003345B} is a simple extension of the algorithm we present in this paper. 

We have also described an algorithm for optimization of the GP kernel parameters given a
measured galaxy ellipticity catalog or a cosmologically-informed prior based on simulation
of lensing fields. An interesting future extension of this work could include both galaxy 
positions and ellipticities in the reconstruction of the gravitational potential. 

Our algorithm is computationally challenging in the solution of linear systems of dimension equal
to the number of galaxies. Next generation surveys will have $10^{7}$--$10^{10}$ galaxies, requiring
both parallelization of the linear solver routines and optimizations such as sparse matrix
approximations via tapering~\citep{Kaufman2008} or Fast Fourier Transform (FFT)-based matrix
multiplications~\citep[exploiting the special structure of the isotropic cosmological covariances][]{padmanabhan++2003}.
As a next step in the validation and scaling of
our code we plan to apply the algorithms in this paper to the variable shear `branches' of the
\greatthree challenge\footnote{\url{http://great3challenge.info/}}~\citep{Mandelbaum++2013}, which
contain $2\times 10^{5}$ galaxies per simulated field. Coincidentally, each of the five
DLS four square degree fields contains a comparable number of $\sim2\times10^5$ lensing source
galaxies.

For the algorithm we validated in this paper we assumed an approximate data vector of galaxy
ellipticity measurements rather than the more informative imaging pixel data for each galaxy, as we
described previously in \citet{mbi-theory}.
If we use galaxy samples rather than an ellipticity catalog then we have to go back to
\autoref{eq:marg_like} and replace the integral over the intrinsic ellipticities with a numerical
importance sampling formula from \citet{mbi-theory}. In this case we do not need to assume
conjugacy of the intrinisc ellipticity prior and the likelihood function, which is good because
the pixel-level likelihood will not be conjugate. The algorithms we presented in
\autoref{sub:cosmo_params} for cosmological parameter estimation are applicable, however, when using
either an approximate ellipticity data vector or the interim sampling of galaxy image model
parameters. 

We can still derive a Wiener Filter from the product of the Gaussian intrinsic ellipticity
distribution (or DP base distribution) and the interim variable shear GP prior assuming a weak
shear approximation. But then we need to evaluate \autoref{eq:posterior_cov} and
\autoref{eq:posterior_mean} for every interim sample.
The parameters of $\smat+\nmat$ include the GP parameters in $\smat$ and the $\alpha$ intrinsic
ellipticity distribution parameter in $\nmat$. If we update either parameter we need to perform a
new matrix factorization and solve operation at every step (but these operations can be done once
for all interim samples of galaxy image model parameters in a given step). Exploring optimized
linear algebra approaches and effective sampling strategies for this more complete framework will
be a focus of future work.

We have not included any redshift information about the source galaxies in our model. However, the 
use of such information (via photometric redshifts) is a critical component of the weak lensing 
analyses for cosmic shear surveys as well as cluster mass reconstructions with optimized 
signal-to-noise ratios. A simple extension of our work to include redshifts could be to impose 
separable GP priors on the lens fields for galaxy samples binned in redshift (or redshift estimator).
The physical correlations between the lens fields inferred in each source bin can be modeled in 
the hierarchical inference stage when marginalizing lens fields to infer cosmological parameters
\citep[as outlined in][]{mbi-theory}. A more thorough approach would be to include probabilistic 
redshifts for each source galaxy and marginalize over each source redshift distribution as 
part of the lens field inference. 

\section*{Acknowledgments}
We thank Chris Paciorek for discussion about the
statistical framework for performing shear inference and the use of Gaussian Processes.
We thank Alex Malz for critical reviews of draft versions of this work.
Thanks to M. James Jee for providing the Deep Lens Survey shear catalog.
Part of this work performed under the auspices of the U.S. Department of Energy
by Lawrence Livermore National Laboratory under Contract DE-AC52-07NA27344.
Funding for this work was provided by LLNL Laboratory Directed Research and Development
grant 16-ERD-013.
This work was also supported by the Director, Office of Science, Office of Advanced
Scientific Computing Research, Applied Mathematics program of the U.S. Department of Energy under
Contract No.DE-AC02-05CH11231.
This research used resources of the National Energy Research Scientific
Computing Center (NERSC), a DOE Office of Science User Facility supported by
the Office of Science of the U.S. Department of Energy under Contract No.
DE-AC02-05CH11231.
This work uses a modified version of the public code \texttt{George} available at
\url{https://github.com/karenyyng/george}, which was forked from
\url{https://github.com/dfm/george}. We also made use of the \texttt{daft} package 
(developed by Dan Foreman-Mackey, David W. Hogg, and contributors, and available at 
\url{http://daft-pgm.org/}) for plotting probabilistic graphical models.

\bibliographystyle{apj3auth}
\bibliography{mbi-references}
\appendix

\section{Gaussian process covariances}
\label{sec:Gaussian process covariances}

We use throughout our analysis the the squared exponential kernel for the GP prior on the
lens potential as specified in \autoref{eq:gp_cov}.
A squared exponential kernel only depends on the
distances between pairs of galaxy locations. This chosen form is invariant
under translational and rotational transformations (homogeneous and isotropic).
Compared with other forms of isotropic covariance kernels,
the exponential squared kernel produces spatial fields that are
smoother for the same correlation length parametrization $\lsqgp = \exp(-1/(8 \ln
\rhogp))$, where $\rhogp$ is a correlation coefficient on the interval $[0,1]$
provided the locations $\xv$ are normalized to the unit interval.
Furthermore, the exponential squared kernel is infinitely differentiable. This
infinitely differentiable kernel choice allows us to derive an analytical expression to
relate different lensing observables through differentiation.

Since the covariance operator is linear,
the derivatives for the covariance kernels of interests are:
\begin{align}
	\label{eq:kernel_derivatives1}
	\Cov (\kappa(\xv), \kappa(\yv))
&= \frac{1}{4}\left(
\kerngp_{,1111} + \kerngp_{,1122} + \kerngp_{,2211} + \kerngp_{,2222}
\right), \\
\Cov(\kappa(\xv), \gamma_1(\yv)) &= \frac{1}{4}\left(
\kerngp_{,1111} + \kerngp_{,2211} - \kerngp_{,1122} - \kerngp_{,2222}
\right), \\
\Cov(\kappa(\xv), \gamma_2(\yv)) &= \frac{1}{4}\left(
\kerngp_{,1112} + \kerngp_{,2212} + \kerngp_{,1121} + \kerngp_{,2221}
\right),\\
\Cov(\gamma_1(\xv), \gamma_1(\yv)) &= \frac{1}{4}\left(
\kerngp_{,1111} - \kerngp_{,1122} - \kerngp_{,2211} + \kerngp_{,2222}
\right), \\
\Cov(\gamma_1(\xv), \gamma_2(\yv)) &= \frac{1}{4}\left(
\kerngp_{,1112} + \kerngp_{,1121} - \kerngp_{,2212} - \kerngp_{,2221}
\right), \\
\Cov(\gamma_2(\xv), \gamma_2(\yv)) &= \frac{1}{4}\left(
\kerngp_{,1212} + \kerngp_{,1221} + \kerngp_{,2112} + \kerngp_{,2121}
\right),
	\label{eq:kernel_derivatives2}
\end{align}
where
\begin{equation}
	\kerngp_{,hijk} = \frac{\partial^4 \kerngp}{\partial x_h \partial x_i \partial y_j \partial y_k},
\end{equation}
and we have defined the shorthand for spatial derivatives with
subscripts for $h,i,j,k = 1, 2$ after a comma.

Generalizing the definition of $s^2$ from \autoref{eq:euclid_dist} to include an arbitrary
distance metric,
\begin{equation}
  s^2 = (\xv-\yv)^T \Matrix{D}(\xv-\yv),
\end{equation}
and with the definition of ${\bf \chi}_i$ as follows:
\begin{equation}
	\frac{\partial s^2}{\partial \xv_i} = -\frac{\partial
	s^2}{\partial \yv_i} =
	2 \Matrix{D}(\xv - \yv)_i \equiv 2{\bf \chi}_i,
\end{equation}
we can show that each entry $\nu_{,hijk}$ of $\kerngp_{,hijk}$ is
related to each entry $\nu$ of the original exponential squared kernel
$\kerngp$ by:

\begin{equation}
  \nu_{,x_h x_i y_j y_k} = (\beta^4 \chi_h \chi_i \chi_j \chi_k -
  \beta^3 (\chi_h \chi_i \Matrix{D}_{jk} \delta_{jk} + 5~{\rm perm.}) + \beta^2
  (\Matrix{D}_{hj} \Matrix{D}_{ik}\delta_{hj}\delta_{ik} + 2~{\rm perm.})) \nu,
  \label{eq:4thderivatives}
\end{equation}
with $\beta \equiv -1/\left(8 \ln \rhogp\right)=\ln(\lsqgp)$. There are 6 permutations (abbreviated as perm.) of the terms in
\ref{eq:4thderivatives}
multiplied by $\beta^3$ due
to choosing two pairs of indices from the $h,i,j,k$, where the order of each
pair of indices matters. Likewise, there are 3 possible permutations of
the terms multiplied by $\beta^2$ due to choosing two pairs of indices from
$h, i, j, k$, where the order of the pairs does not matter.
The full derivation of the relevant derivatives and terms can be found in
\citet{ng2016}. The full covariance model is implemented in a fork of the 
public GP code 
\texttt{George}\footnote{\url{https://github.com/dfm/george}\\ 
\url{https://github.com/karenyyng/george}}~\citep{ambikasaran2013HODLR}.

\begin{figure}[!htb]
  \centerline{
    \includegraphics[width=0.6\linewidth]{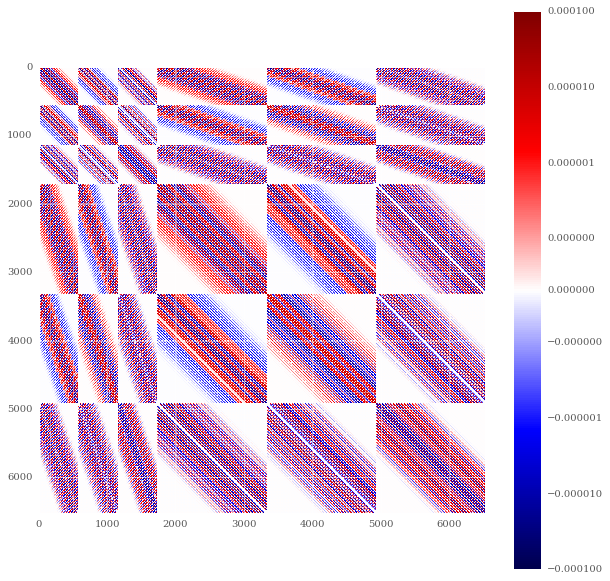}
  }
  \caption{The Gaussian Process covariance used in our simulation validation study.
  This covariance matches that used for \autoref{fig:mass_map_comparison_galsim}, corresponding
  to $\sigma_e = 0.0026$. There are 1600 galaxy locations, 576 interpolation grid locations,
  and 3 lens fields for a total covariance dimension of $6528^2$.
  There are 4 large sub-blocks per dimension. The first large sub-block represents the grid
  locations to which we interpolate from the galaxy locations. The 3 remaining large sub-blocks
  represent $\kappa, \gamma_1, \gamma_2$ at the galaxy locations. Inside the first large sub-block
  is the $\kappa, \gamma_1, \gamma_2$ sub-sub-blocks for the grid locations.
  }
  \label{fig:gp_cov}
\end{figure}

We show the structure of the GP covariance matrix, $\kerngp$, from \autoref{eq:gp_cov} in
\autoref{fig:gp_cov} for a simulation with 1600 galaxies placed on a uniform $40\times40$ grid on
the sky and a coarser grid of $24\times24$ locations to which the lens fields are interpolated.
The distinctive banded structure is because of the galaxy locations on a uniform 2D grid that is
then flattened into a 1D array. The sub-blocks correspond to the $\kappa, \gamma_1, \gamma_2$ fields
for the grid and galaxy locations. A notable feature of the covariance in \autoref{fig:gp_cov} is
the rapid decrease in absolute values of the elements moving away from the diagonals of each
sub-block. We will explore in future work whether sparse linear algebra software packages may
be useful for speeding up the computations in our lens field inference.

It is possible to generalize the derivation of derivatives by using numerical
differentiation. Numerical differentiation may allow evaluation of other
isotropic kernels, such as the Mat\'{e}rn kernels.
An implementation via numerical differentiation  may
or may not be more computationally intensive to
compute, depending on the required numerical precision (aka order of accuracy).
Alternative differentiation technology such as automatic differentiation can be
considered but is outside the scope of this paper.

\section{Cosmology-dependent covariance model} 
\label{sec:cosmology_dependent_covariance_model}

Under linear cosmological perturbation theory, we can model the lensing convergence as Gaussian
distributed with a specified isotropic angular power spectrum. This power spectrum then defines a
cosmological covariance model for the lens fields, which we derive and explore in this section.

\begin{figure}[!htb]
    \centerline{
        \includegraphics[width=0.41\textwidth]{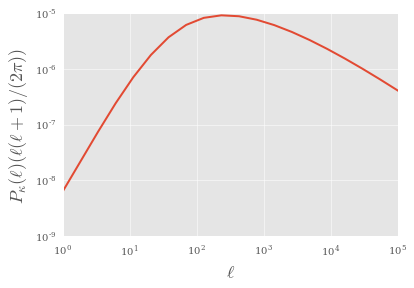}
        \includegraphics[width=0.40\textwidth]{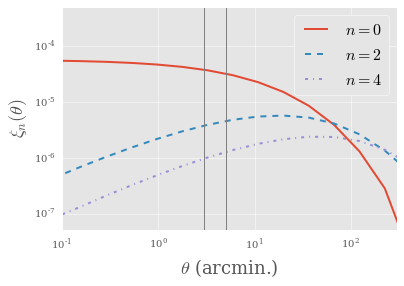}
    }
    \caption{
    Left: The convergence power spectrum used to simulate shear fields in our simulation study in
    \autoref{sub:simulation_study}. The power spectrum is calculated for linear cosmological
    perturbations with a source distribution narrowly peaked at $z=1$.
    Right: The angular correlation functions obtained by integrating the power spectrum from the
    left panel with different Bessel function weights. We perform the Bessel function integrals
    using the Hankel transforms implemented in~\textsc{FFTLog}~\citep{fftlog}.
    The vertical lines denote the grid sizes for the input galaxies and the interpolation grid
    in \autoref{sub:simulation_study}. Angular scales to the left of these lines are unresolved in
    our simulation grids.
    }
    \label{fig:cosmo_corr}
\end{figure}

Gravitational lensing generates only E-mode shear distortions for a uniform distribution of
source illumination. Clustered sources can generate B-mode shear, but only on sub-arcminute
scales for typical cosmological models~\citep{2002A&A...389..729S}. We consider then a real-valued
lensing convergence $\kappa = \kappa_{E}$ with an associated angular power spectrum $P_{E}(\ell)$,
under a flat-sky approximation.
In the left panel of \autoref{fig:cosmo_corr} we show the angular convergence power spectrum used
in our simulation study from \autoref{sub:simulation_study}.

The covariance of shear and convergence is most easily described in terms of the rotated shear
components~\citep{schneider2002,joachimi08},
\begin{equation}
  \shear_{t} \equiv -{\rm Re}\left(\shear e^{-2i\phi} \right)
  \quad
  \shear_{\times} \equiv -{\rm Im}\left(\shear e^{-2i\phi}\right),
\end{equation}
where $\phi$ is the polar angle of the angular separation vector of a galaxy pair.
In the flat-sky limit, the two-point functions of the convergence and shear components are then
related to the convergence power spectrum
as~\citep[in analogy with CMB polarization correlation functions][]{kamionkowsky1997},
\begin{align}
  \left<\kappa_E\kappa_E\right>(|\xv|) &=
    \int_{0}^{\infty} \frac{\ell d\ell}{2\pi}\,
    P_{E}(\ell) J_{0}(\ell|\xv|)
  \\
  \left<\kappa_E\shear_t\right>(|\xv|)  &=
    \int_{0}^{\infty} \frac{\ell d\ell}{2\pi}\,
    P_{E}(\ell) J_{2}(\ell|\xv|)
  \\
  \left<\shear_t\shear_t\right>(|\xv|) &=
    \half  \int_{0}^{\infty} \frac{\ell d\ell}{2\pi}\,
    P_{E}(\ell) \left[J_{0}(\ell|\xv|) + J_{4}(\ell|\xv|)\right]
  \\
  \left<\shear_{\times}\shear_{\times}\right>(|\xv|) &=
    \half \int_{0}^{\infty} \frac{\ell d\ell}{2\pi}\,
    P_{E}(\ell)  \left[J_{0}(\ell|\xv|) - J_{4}(\ell|\xv|)\right]
    ,
\end{align}
where $J_{n}$ is the $n$-th order Bessel function of the first kind. We can recover the
correlation functions of the shear components measured with respect to a fixed sky coordinate
system from,
\begin{align}
  \shear_1 &= -\cos(2\phi)\shear_{t} + \sin(2\phi)\shear_{\times}
  \notag\\
  \shear_2 &= -\sin(2\phi)\shear_{t} - \cos(2\phi)\shear_{\times}.
\end{align}
Then,
\begin{align}
  \left<\shear_1\shear_1\right> &= \cos^2(2\phi)\left<\shear_t\shear_t\right>
    + \sin^2(2\phi)\left<\shear_{\times}\shear_{\times}\right>
  \notag\\
  \left<\shear_2\shear_2\right> &= \sin^2(2\phi)\left<\shear_t\shear_t\right>
    + \cos^2(2\phi)\left<\shear_{\times}\shear_{\times}\right>
  \notag\\
  \left<\shear_1\shear_2\right> &= \sin(2\phi)\cos(2\phi)
    \left(\left<\shear_t\shear_t\right> - \left<\shear_{\times}\shear_{\times}\right>\right).
\end{align}
We show the angular correlation functions contributing to the convergence and shear auto-
and cross-correlations in the right panel of \autoref{fig:cosmo_corr}. Specifically we
plot,
\begin{equation}
  \xi_{n}(\theta) \equiv \int \frac{\ell d\ell}{2\pi}\, P_{E}(\ell) J_{n}(\ell\theta).
\end{equation}

\begin{figure}[!htb]
  \centerline{
    \includegraphics[width=0.6\linewidth]{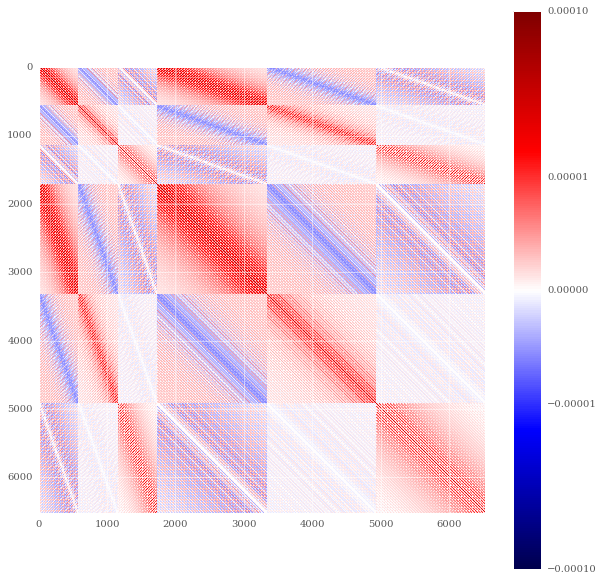}
  }
  \caption{The covariance of convergence and shear derived from the cosmological power spectrum used in our simulation validation study.
  }
  \label{fig:cosmo_cov}
\end{figure}

We show the resulting joint covariance of convergence and shear in \autoref{fig:cosmo_cov}. The
covariance is evaluated at the same sky locations as in \autoref{fig:gp_cov}. It is notable that
the covariance in \autoref{fig:cosmo_cov} has larger relative amplitudes between the diagonal and
off-diagonal components of each sub-block than in \autoref{fig:gp_cov}. This indicates that the
effective kernel for averaging shears to infer the convergence is broader under this cosmological
model than in the covariance we determined under the GP model. Our GP model is thus discarding more
information from distant galaxies in inferring the convergence at any given sky location.

\begin{figure*}[!htb]
  \centerline{
    \includegraphics[width=0.54\textwidth]{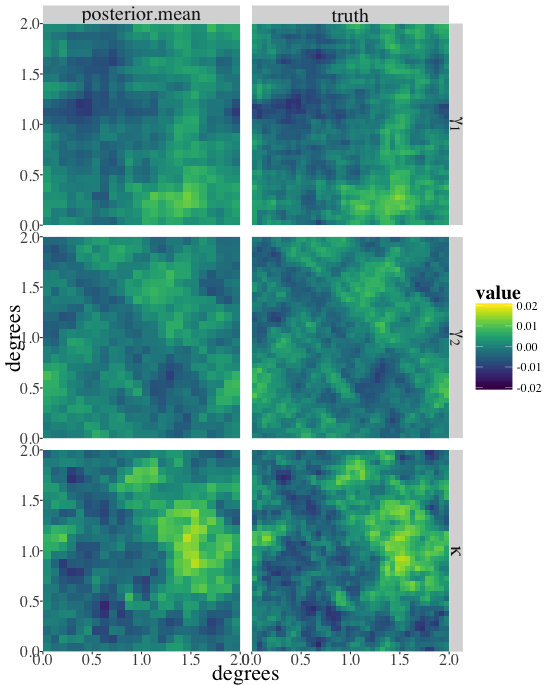}
    \includegraphics[width=0.35\textwidth]{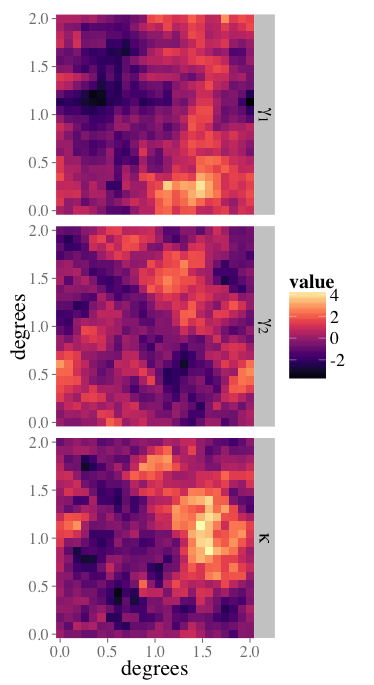}
  }
  \caption{Left: Comparison of the convergence and shear maps in our simulation study between that
  used to generate our mock galaxy ellipticity catalog (right column) and the output of our
  GP interpolation (left column). The rows show the maps for the two shear components $\gamma_{1,2}$
  and the convergence $\kappa$. These maps cover a $2\times2$ square degree field. The simulated
  intrinsic ellipticity r.m.s.\ is set to an artificially small value of $\sigma_e=0.0026$,
  which is 100 times smaller than that observed for the complete Deep Lens Survey catalog.
  Right: signal-to-noise ratio (SNR) maps for the same simulations. We calculate SNR as the
  ratio of the absolute value of the map to the square root of the diagonal of the covariance
  in \autoref{eq:posterior_cov}.
  These simulations use 1600 galaxies and grid sizes for the GP interpolated shear and convergence
  with 24 grid cells per dimension.
  }
  \label{fig:mass_map_comparison_galsim_cosmo}
\end{figure*}

To further understand the relative performance of the cosmological and GP model covariances for
lens field mapping, in \autoref{fig:mass_map_comparison_galsim_cosmo} we reproduce the maps
shown in \autoref{fig:mass_map_comparison_galsim} but now using the cosmology covariance instead of
the GP kernel. The reconstruction of all lens fields in the left panel of
\autoref{fig:mass_map_comparison_galsim_cosmo} is superb given the different input and interpolated
grid sizes. This is expected because the input data is strictly Gaussian distributed and we used
the same power spectrum to simulate the data and to perform the WF interpolation.
However, comparing the right panel of \autoref{fig:mass_map_comparison_galsim_cosmo} with
that of \autoref{fig:mass_map_comparison_galsim} shows the SNR of the mean posterior maps
using the cosmology covariance is smaller than that using the GP covariance. This is because the
WF variance using the cosmology covariance is larger than that using the GP covariance.

\begin{figure}[!htb]
  \centerline{
    \includegraphics[width=0.5\textwidth]{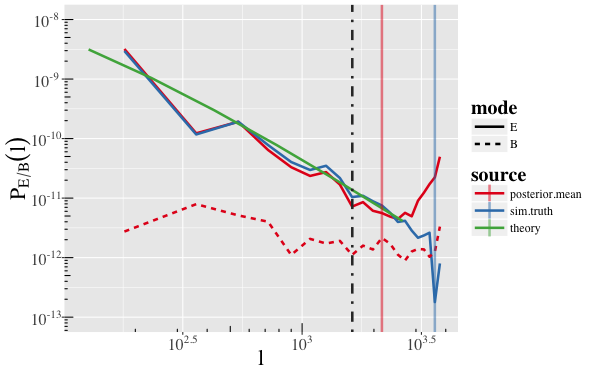}
  }
  \caption{Comparison of power spectrum estimators in our simulation study to the `truth' input
  spectrum.
  The vertical solid lines show the Nyquist frequencies of the grids.
  The vertical dot-dashed matches that in \autoref{fig:power_spectra_galsim} to aide
  comparison of the figures, but has no relationship to the plotted power spectra here.
  }
  \label{fig:power_spectra_galsim_cosmo}
\end{figure}

\begin{figure*}
  \centerline{
    \includegraphics[width=0.6\textwidth]{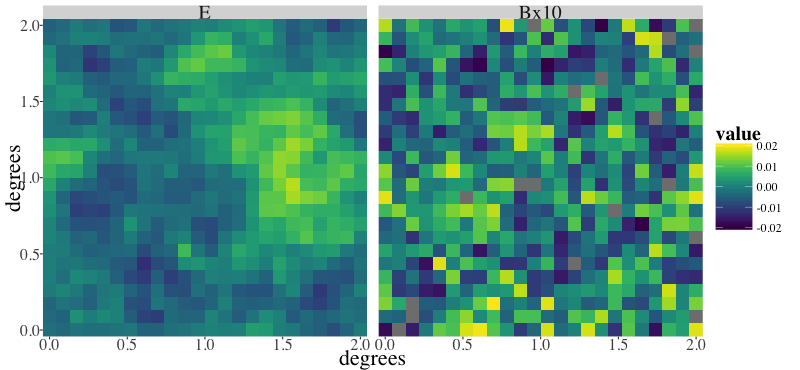}
  }
  \caption{E and B mode maps derived from the posterior mean shear maps shown in the left column
  of \autoref{fig:mass_map_comparison_galsim_cosmo}.
  Note the value of the plotted B-mode map
  has been multiplied by 10 for easier visualization.
  }
  \label{fig:EB_maps_cosmo}
\end{figure*}

In \autoref{fig:power_spectra_galsim_cosmo} we show the E and B mode power spectra derived from the
maps in \autoref{fig:mass_map_comparison_galsim_cosmo}. The E-mode power spectrum reconstruction
is more accurate at frequencies closer to Nyquist than that in \autoref{fig:power_spectra_galsim}
derived from the GP model. However, the B-mode power spectrum amplitude is larger in
\autoref{fig:power_spectra_galsim_cosmo} than in \autoref{fig:power_spectra_galsim}.
We show the corresponding E and B mode maps derived from the mean posterior using the cosmology
covariance in \autoref{fig:EB_maps_cosmo}. The E-mode map in \autoref{fig:EB_maps_cosmo} is
an even better match to the input convergence field than the posterior convergence map in
\autoref{fig:mass_map_comparison_galsim_cosmo}. This indicates there is some level of B-mode
leakage in the posterior convergence map in \autoref{fig:mass_map_comparison_galsim_cosmo}.
The B-mode map in \autoref{fig:EB_maps_cosmo} has non-negligible amplitude
(consistent with \autoref{fig:power_spectra_galsim_cosmo}) and contains more spatial structure
than the B-mode map in \autoref{fig:EB_maps} derived from the GP prior.

In summary, the comparison of the mean posterior lens fields and corresponding power spectra between
the GP and cosmological covariance models shows the GP covariance yields higher SNR mass map extrema
and smaller B-mode power at the expense of reduced dynamic range in the power spectrum relative to
the cosmological model covariance (all conditioned on the particular choice of GP parameters).

\section{Linear theory cosmological parameter dependence} 
\label{sec:linear_theory_cosmological_parameter_dependence}

In this section we derive the form of the covariance that appears in the
marginal likelihood for galaxy ellipticities when assuming linear theory,
weak shear, and a Gaussian likelihood as in \autoref{eq:marg_post_cosmo_linear}.

The marginal likelihood requires evaluation of \autoref{eq:lens_field_cond_factor}, which
has as left-hand side,
\begin{equation}\label{eq:ltc1}
  \prf{\lensparams | \lensparams', \gpparams} \prf{\lensparams' | \theta}.
\end{equation}
Each term in \autoref{eq:ltc1} is a Normal distribution, so evaluating the product
in \autoref{eq:ltc1} reduces to evaluating the sum of the inverse covariances of
each term. It is convenient to evaluate the identical expression to \autoref{eq:ltc1},
\begin{equation}
  \prf{\lensparams | \lensparams', \gpparams} \prf{\lensparams' | \theta}
  =
  \prf{\lensparams | \lensparams', \gpparams} \prf{\lensparams' | \gpparams}
  \frac{\prf{\lensparams' | \theta}}{\prf{\lensparams' | \gpparams}}.
\end{equation}
Next, factor the GP covariance in $\prf{\lensparams | \lensparams', \gpparams} \prf{\lensparams' | \gpparams}$
into blocks that depend on coordinates $\xv$ and $\xv'$,
\begin{equation}\label{eq:gp_cov_blocks}
  \smatgp = \left(
  \begin{array}{cc}
    \smatgp(\xv,\xv) & \smatgp(\xv,\xv') \\
    \smatgp(\xv',\xv) & \smatgp(\xv',\xv')
  \end{array}
  \right)
  \equiv \left(
  \begin{array}{cc}
    \amat & \bmat \\
    \bmat^{T} & \cmat
  \end{array}
  \right),
\end{equation}
where the second equality defines a short-hand notation for the matrix blocks.
We can invert \autoref{eq:gp_cov_blocks} using the expressions for the
conditional covariance, or Schur complement~\citep[e.g.,][]{COTTLE1974189},
\begin{equation}\label{eq:gp_cov_block_inverse}
  \smatgp^{-1} = \left(
  \begin{array}{cc}
    \wmat & -\wmat\bmat\cmat^{-1} \\
    -\cmat^{-1}\bmat^{T}\wmat & \cmat^{-1} + \cmat^{-1}\bmat^{T}\wmat\bmat\cmat^{-1}
  \end{array}
  \right)
\end{equation}
where,
\begin{equation}
  \wmat \equiv \left(
    \amat - \bmat\cmat^{-1}\bmat^{T}
  \right)^{-1}.
\end{equation}
Subtracting the covariance for $\prf{\lensparams' | \gpparams}$ from \autoref{eq:gp_cov_block_inverse}
and adding the covariance $\Sigmamat(\theta)$ for $\prf{\lensparams' | \theta}$ yields the
covariance,
\begin{equation}
  \mmat^{-1} \equiv \left(
  \begin{array}{cc}
    \wmat & -\wmat\bmat\cmat^{-1} \\
    -\cmat^{-1}\bmat^{T}\wmat & \Sigmamat^{-1}(\theta) + \cmat^{-1}\bmat^{T}\wmat\bmat\cmat^{-1}
  \end{array}
  \right).
\end{equation}
Inverting gives the desired covariance matrix with a Gaussian cosmology prior imposed on the
lens fields $\lensparams(\xv')$ at locations $\xv'$,
\begin{equation}\label{eq:signal_cov_cosmo}
    \Sigmamat(\xv, \xv'; \gpparams, \theta) = \left(
    \begin{array}{cc}
      \amat - \bmat\cmat^{-1}\bmat^{T} + \bmat\cmat^{-1}\Sigmamat(\theta)\cmat^{-1}\bmat^{T}
        & \bmat\cmat^{-1}\Sigmamat(\theta) \\
      \Sigmamat(\theta)\cmat^{-1}\bmat^{T}
        & \Sigmamat(\theta)
    \end{array}
    \right).
\end{equation}
As expected, \autoref{eq:signal_cov_cosmo} reduces to $\Sigmamat(\xv';\theta)$ when the
lens fields are evaluated only at the locations of the theory predictions $\xv'$.

\end{document}

%% file: mbi-general-macros.tex
\def \spose#1{\hbox  to 0pt{#1\hss}}  
\newcommand {\lta} {\mathrel{\spose{\lower 3pt\hbox{$\sim$}}\raise  2.0pt\hbox{$<$}}}
\newcommand {\gta} {\mathrel{\spose{\lower  3pt\hbox{$\sim$}}\raise 2.0pt\hbox{$>$}}}

\newcommand {\kms} {\ifmmode  \,\rm km\,s^{-1} \else $\,\rm km\,s^{-1}  $ \fi }
\newcommand {\kpc} {\ifmmode  {\rm kpc}  \else ${\rm  kpc}$ \fi  }  
\newcommand {\pc} {\ifmmode  {\rm pc}  \else ${\rm pc}$ \fi  }  
\newcommand {\Msun} {\ifmmode {\rm M_{\odot}} \else ${\rm M_{\odot}}$ \fi} 
\newcommand {\Zsun} {\ifmmode {\rm Z_{\odot}} \else ${\rm Z_{\odot}}$ \fi} 
\newcommand {\yr} {\ifmmode yr^{-1} \else $yr^{-1}$ \fi} 
\newcommand {\hMsun} {\ifmmode h^{-1}\,\rm M_{\odot} \else $h^{-1}\,\rm M_{\odot}$ \fi}

\usepackage[usenames]{color}


%% file: mbi-notation-macros.tex
\newcommand{\half}{\frac{1}{2}}

\newcommand{\ident}{\mathbb{1}} 

\newcommand{\xv}{\mathbf{x}}
\newcommand{\yv}{\mathbf{y}}

\newcommand{\muv}{{\bm \mu}}

\newcommand{\psf}{{\Pi}}

\newcommand{\ngal}{n_{\rm gal}}

\newcommand{\shear}{\gamma}
\newcommand{\normdist}{\mathcal{N}}

\newcommand{\gravpotfinal}{\Psi^{\rm LT}}
\newcommand{\gravpotinit}{\Psi^{\rm IC}}
\newcommand{\lenspot}{\psi}
\newcommand{\lensparams}{\Upsilon}

\newcommand{\Sigmamat}{\mathsf{\Sigma}}
\newcommand{\covinit}{\mathsf{\Sigma}^{\rm IC}}

\newcommand{\smat}{\mathsf{S}}
\newcommand{\smatgp}{\smat_{\rm GP}}
\newcommand{\mmat}{\mathsf{M}}
\newcommand{\nmat}{\mathsf{N}}
\newcommand{\amat}{\mathsf{A}}
\newcommand{\bmat}{\mathsf{B}}
\newcommand{\cmat}{\mathsf{C}}
\newcommand{\wmat}{\mathsf{W}}

\newcommand{\kerngp}{\mathsf{S}}
\newcommand{\rhogp}{\rho_{\rm GP}}
\newcommand{\lambdagp}{\lambda_{\rm GP}}
\newcommand{\lsqgp}{\ell^{2}_{\rm GP}}

\newcommand{\gpparams}{a_{\rm GP}}

\def\greatthree{{\sc GREAT3}\xspace}

\newcommand{\chomp}{{\sc CHOMP}\xspace}
\newcommand{\galsim}{{\sc GalSim}\xspace}
\newcommand{\lensent}{{\sc LensEnt2}\xspace}

\def\pr{{\rm Pr}}
\newcommand{\prf}[1]{\pr\left(#1\right)}
\def\data{{\mathbf{d}}}

\def\eint{\epsilon^\mathrm{int}}

%% file: mbi-addresses.tex
\def\kipac{Kavli Institute for Particle Astrophysics and Cosmology, Stanford University, 
Stanford, CA 94035, USA.
}
\def\slac{SLAC National Accelerator Laboratory, 
Menlo Park, CA 94025, USA.}
\def\ccin2p3{Centre de Calcul de l'IN2P3, USR 6402 du CNRS-IN2P3, 43 Bd. du 11 Novembre 1918, 69622 Villeurbanne Cedex, France.} 

\def\llnl{Lawrence Livermore National Laboratory, 
Livermore, CA 94551, USA.}
\def\davis{University of California, Davis, 
Davis, CA 95616, USA.}

\def\NERSC{National Energy Research Scientific Computing Center, Lawrence
Berkeley National Laboratory, 1 Cyclotron Road, Berkeley, CA 94720-8150, USA.}